\documentclass{article}
\usepackage[utf8]{inputenc}
\usepackage{authblk}
\usepackage{setspace}
\usepackage[margin=1.25in]{geometry}
\usepackage{graphicx}
\graphicspath{ {./figures/} }
\usepackage{subcaption}
\usepackage{amsmath,amsthm,amsfonts,amssymb,amscd,epsfig,enumerate,yfonts}
\usepackage{tikz}
\usepackage{float}
\usetikzlibrary{positioning}
\usepackage{booktabs}
\usepackage{ragged2e}
\usepackage{hyperref}
\hypersetup{
    colorlinks=true,
    linkcolor=blue,
    filecolor=blue,      
    urlcolor=blue,
    }
\title{Periodic orbits and their gravitational wave radiations around the Schwarzschild-MOG black hole}
\author[1]{Oreeda Shabbir\thanks{Email: oreedashabbir7@gmail.com}}
\author[2]{Mubasher Jamil\thanks{Email: mjamil@sns.nust.edu.pk}}
\author[3]{Mustapha Azreg-A\"{\i}nou\thanks{Email: azreg@baskent.edu.tr}}
\affil[1,2]{School of Natural Sciences, National University of Sciences and Technology (NUST), H-12, Islamabad (44000), Pakistan}
\affil[3]{Ba\c{s}kent University, Engineering Faculty, Ba\u{g}l\i ca Campus, 06790-Ankara, T\"{u}rkiye}

\date{}

\begin{document}
\maketitle
\begin{abstract}
    This article explores the motion of massive particles in the gravitational field of a modified gravity (MOG) black hole (BH), characterized by the parameter $\alpha$. Using the Hamiltonian formalism, the geodesic equations and the effective potential governing particle trajectories are derived. Key features, including the innermost stable circular orbit (ISCO) and the innermost bound circular orbit (IBCO), are analyzed, revealing their dependence on the particle’s energy, angular momentum, and the MOG parameter. In the extremal case, where $\alpha=-1$, the event horizon merges with the Cauchy horizon, forming a distinctive BH configuration. Numerical methods are employed to compute periodic orbits in this spacetime, with a comparison drawn to the Schwarzschild BH. The findings indicate that for $\alpha>0$, periodic orbits around Schwarzschild-MOG BH exhibit lower energy requirements than those in Schwarzschild spacetime, whereas for $-1<\alpha<0$, the energy requirements are higher. Precessing orbits near periodic trajectories are also examined, offering insights into their complex dynamical behavior. Finally, the gravitational wave (GW) radiation from the periodic orbits of a test particle around the Schwarzschild-MOG BH is examined, generating intricate waveforms that provide insights into the gravitational structure of the system.
\end{abstract}
\begin{justify}
\section{Introduction}
BHs are incredibly dense celestial bodies with intense gravitational attraction, anticipated by general relativity (GR). BHs gained significant attention when Karl Schwarzschild found the first exact solution of Einstein field equations in 1916 \cite{schwarzschild1916gravitationsfeld}. BHs give rise to various astronomical phenomena, such as GWs \cite{jia2023gravitational}, oscillations in spacetime caused by accelerating masses \cite{andersson2002oscillations}, gravitational lensing \cite{bodenner2003deflection}, shadows \cite{vagnozzi2019hunting}, etc. In recent years, various astronomical observations of the dynamics of the entire universe and galaxies have been made, providing evidence for the existence of influential forms of matter, namely dark energy and dark matter \cite{bertone2018history,bertone2018new}. Dark matter has never been directly observed. It comprises most of the matter in the universe, about six times more abundant than ordinary visible matter, and behaves as the unseen scaffolding on which galaxies form and develop. It particularly emits no radiation and interacts only with the action of gravity. Over the years, physicists have observed a variety of dark matter, including hypothetical particles such as sterile neutrinos, “weakly interacting massive particles" (WIMPS), and axions \cite{freeman2006exploring}. One important astronomical system that is a focus of dark matter studies is the galactic scale dynamics \cite{bertone2005particle}. However, a complete theoretical framework to describe the behavior and nature of this dark matter is still lacking.\\ \indent GR also faces some fundamental challenges that contradict quantum field theory (QFT), such as the singularities in certain solutions. These singularities pose significant challenges for reconciling GR with QFT, as QFT successfully describes three of the four fundamental forces but struggles to incorporate gravity within its framework. Researchers have proposed alternative theories of gravity or modifications to GR to address this issue. Since Einstein's theory of GR has been modified in various ways, one such modification is scalar-tensor-vector gravity (STVG), proposed by Moffat \cite{moffat2006scalar,moffat2006time}. This theory is also known as MOG. It postulates the existence of a massive vector field and aims to construct a unified theory of gravity.\\ \indent STVG extends GR by introducing scalar, tensor, and massive vector fields to modify the gravitational interaction \cite{moffat2006scalar}. These fields enable the theory to dynamically adjust the gravitational constant $G$ based on local spacetime conditions, accounting for variations in gravitational behavior across different cosmic scales. This modification offers a more flexible framework for explaining phenomena such as rotation curves of galaxies \cite{moffat2013mog,moffat2015rotational}, the dynamics of galactic clusters \cite{moffat2014mog}, and cosmological observations, without the need for dark matter, making STVG a promising alternative to traditional GR. The STVG or MOG theory has successfully explained the solar system observations \cite{moffat2014scalar}, the cosmic microwave background (CMB) acoustic spectrum data \cite{moffat2014structure}, and the matter power spectrum. A detailed study of MOG BHs and interior solutions describing compact objects is provided in \cite{moffat2015rotational,moffat2015black}. Additionally, the observable characteristics of BH shadows in MOG have been analyzed, revealing deviations from GR that offer potential observational tests for the theory \cite{moffat2015modified}. Furthermore, the Kerr-MOG-(A)dS BH solution has been studied, offering insights into how STVG influences rotating BHs in asymptotically (Anti-)de Sitter spacetimes \cite{liu2024kerr}.\\
\indent Other MOG theories have been proposed to explain deviations from GR on astrophysical and cosmological scales. Notable models include quintessence fields, string clouds \cite{mustafa2021radial}, and bumblebee gravity \cite{mustafa2025epicyclic}, each addressing distinct phenomena that challenge standard gravity. The impact of quintessence and string clouds on particle motion near BHs reveals significant deviations in orbital behavior, while bumblebee gravity, which incorporates spontaneous Lorentz symmetry breaking, affects epicyclic oscillations around rotating charged BHs. Observational advances, such as the Event Horizon Telescope’s (EHT) image of Sagittarius A*, provide crucial tests of GR and alternative gravity theories at horizon scales \cite{vagnozzi2023horizon}, offering new insights into BH physics and the nature of gravity in extreme environments.\\
\indent The study of geodesics for test particles near BHs provides deep insights into the fundamental phenomena of GR and the nature of gravity. In BH spacetimes, stars can be treated as particles following timelike geodesics near massive BHs, and their motion plays a crucial role in understanding BH physics, detecting gravitational fields, and measuring BH masses. Observations, such as those by Abuter et al. \cite{abuter2020detection}, have confirmed the existence of Schwarzschild precession in the orbits around Sgr A* near the center of the Milky Way, reinforcing the accuracy of GR. In static, spherically symmetric BH spacetimes, Kostić \cite{kostic2012analytical} provided analytical solutions for timelike geodesics, detailing several types of orbit, including scattering, plunging, near, and bound orbits, while Cruz et al. \cite{cruz2005geodesic} explored analytical geodesics around Schwarzschild-AdS BHs.\\ \indent One important type of bound orbit is the periodic orbit, which is particularly significant in GR as it captures essential information about orbits around BHs, with all generic orbits being small deviations from these. Periodic orbits are crucial in addressing complex problems in astrodynamics, such as the motion of planetary satellites and the stability of galactic potentials \cite{lake2004galactic}. Despite their extensive study in classical systems, their behavior in relativistic astrophysical contexts, like compact binary stars or extreme mass-ratio inspirals (EMRI), remains an area of active research \cite{wang2019bright}. Inspired by this concept, Levin and Perez-Giz recently developed a taxonomy of periodic orbits \cite{levin2008periodic}. This taxonomy establishes a fascinating relationship between a triplet of integers $(z,w,v)$ and periodic orbits. These three integers define a rational number $q$, which corresponds to a specific periodic orbit characterized by a given angular momentum and energy of the particle. Using this taxonomy, periodic orbits have been studied around various BH spacetimes, including Schwarzschild \cite{levin2008periodic}, Kerr \cite{levin2009energy}, RN BHs \cite{misra2010rational}, Kerr-Sen BH \cite{liu2019periodic}, as well as naked singularities \cite{babar2017periodic}, etc.
\\ \indent Periodic orbits in binary systems are fundamental to the generation of GWs, particularly during the inspiral and merger phases. Stellar-mass BHs can form tightly bound orbits around significantly larger supermassive BHs, effectively behaving as test particles in these extreme gravitational fields. Such systems, known as EMRIs, are key targets for future space-based GW detectors like LISA \cite{k1997lisa,amaro2017laser}, Taiji \cite{sirunyan2018search}, and Tianqin \cite{luo2016tianqin,gong2021concepts}. The potential detection of these systems offers exciting opportunities to explore the fundamental nature of gravity and cosmology \cite{arun2022new, auclair2023cosmology}. The bound orbits of stellar-mass BHs around a supermassive BH may exhibit peculiar behaviors during the inspiral stage of GW detection, as GW radiation causes them to move closer to the larger BH. During this process, periodic orbits act as continuous transitions, significantly contributing to our understanding of the GW signals emitted throughout the inspiral process and enhancing our broader comprehension of the dynamics of compact objects in extreme gravitational environments \cite{glampedakis2002zoom}.\\
\indent The Schwarzschild-MOG BH geometry offers a compelling framework for exploring deviations from GR within MOG theories. By introducing the MOG parameter $\alpha$, this spacetime allows us to explore the effects of enhanced gravitational interactions on the motion of particles, BH behavior, and GW emission. Such studies are crucial for understanding how MOG influences compact objects and for interpreting observational data related to gravitational signals. Periodic orbits around Schwarzschild-MOG BH offer valuable insights into how $\alpha$ alters the motion of particles. As $\alpha$ increases, the gravitational field becomes more intense \cite{hu2022observational}, enhancing the stability of bound orbits by deepening the effective potential well, which makes it more likely for particles to remain in stable, periodic paths around the BH. However, this stronger gravitational influence also imposes limits on orbital stability, especially near the event horizon. Studying these effects helps to understand how deviations from GR influence orbital dynamics around compact objects \cite{dedeo2004testing}.\\
\indent This paper is organized as follows. In Section 2, we provide a brief review of the Schwarzschild-MOG BH. In Section 3, we derive the geodesic equations using the Hamiltonian formulation. Also, we analyze the behavior of effective potential and circular orbits as well as the derivation of the ISCO and IBCO. In Section 4, we study the periodic orbits around the Schwarzschild-MOG BH. Section 5 encompasses the analysis of processing orbits near periodic orbits. In Section 6, we investigate the GW radiation emitted by periodic orbits. Finally, the discussion and conclusion are given in Section 7. Throughout the paper, we use a geometrized unit system where $G_N=c=1$, along with the metric convention $(-,+,+,+)$.
\section{Schwarzschild-MOG Black Hole}
The Schwarzschild-MOG BH is a static, spherically symmetric vacuum solution. The Schwarzschild-MOG metric in spherical coordinates $(t,r,\theta,\phi)$ is given as \cite{moffat2015black}
\begin{align}
    \label{eq1} ds^2=-\left(1-\frac{2GM}{r}+\frac{GQ^2}{r^2}\right)dt^2+\left(1-\frac{2GM}{r}+\frac{GQ^2}{r^2}\right)^{-1}dr^2+r^2 d\Omega^2.
\end{align}
This solution describes the final stage
of the collapse of a massive, compact object such as a star in terms of an enhanced gravitational constant $G=G_N\left(1+\alpha\right)$, a gravitational repulsive
force with a charge $Q=\sqrt{\alpha G_N}M$, where $\alpha$ is MOG parameter. Here, $G_N$ is Newton's gravitational constant and $M$ is the total mass of the BH. The parameter $\alpha$ defines the strength of the gravitational field: the larger the $\alpha$, the stronger the gravitational field strength. For $\alpha>0$, the Schwarzschild-MOG BH has two positive real roots satisfied by 
 condition $g^{rr}=0$, the largest and smallest roots correspond to the event horizon $(r_e)$ and Cauchy horizon $(r_c)$, respectively. The horizons are \cite{sharif2018neutral}
\begin{align}
r_{e} &=G_\text{N}M\left(1+\alpha+\left(1+\alpha\right)^{\frac{1}{2}}\right),\\
r_{c} &=G_\text{N}M\left(1+\alpha-\left(1+\alpha\right)^{\frac{1}{2}}\right).
\end{align}
In the limit $\alpha=0$, this spacetime reduces to the Schwarzschild BH, and the event horizon is $r_e=2G_{\text{N}}M$. For convenience, put the values of $Q$ and $G$ in Eq.~\eqref{eq1}, then the Schwarzschild-MOG metric takes the form
\begin{align}
    ds^2 &=-fdt^2+\frac{dr^2}{f}+r^2\left(d\theta^2+\sin^2\theta d\phi^2\right),
    \end{align}
    where
    \begin{align}
    f &=1-\frac{2M\left(1+\alpha\right)}{r}+\frac{M^2\alpha\left(1+\alpha\right)}{r^2},~~(-1\leq\alpha<\infty).
\end{align}
 \begin{figure}[H]
 \centering
{{\includegraphics[height=9.5cm,width=13.3cm]{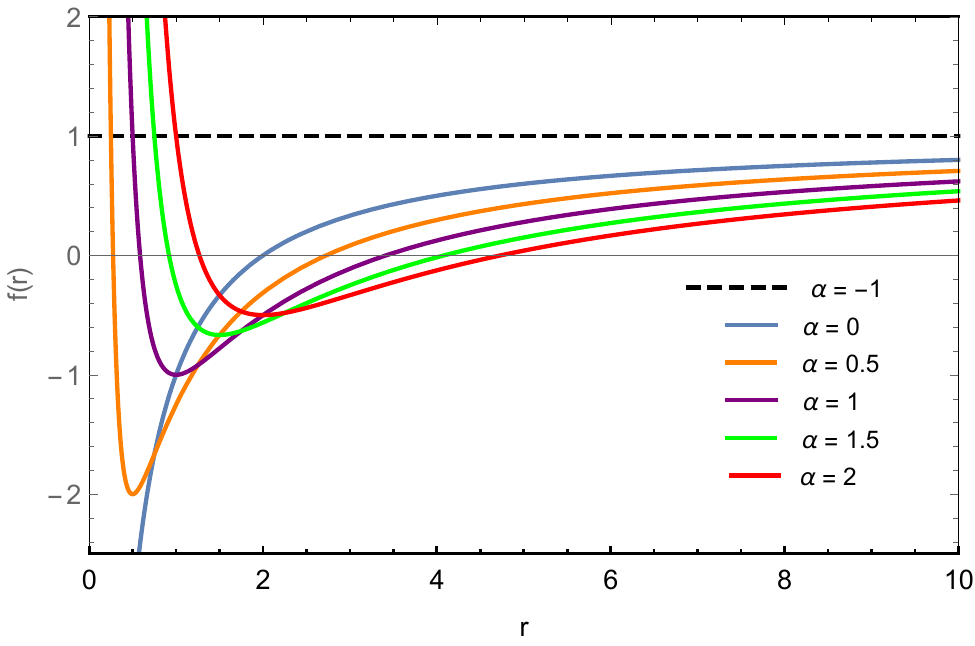}}}
\caption{Behavior of $f(r)-r$ curves for different $\alpha$ values and $M=1$, illustrating the shift between the event and Cauchy horizons.}
\label{fig 3.1}
 \end{figure}
 \begin{justify}
Fig.~\ref{fig 3.1} shows the behavior of the $f(r)$ for different values of $\alpha$. We observe that by increasing $\alpha$ the event horizon drifts farther away from the Cauchy horizon. For $\alpha=-1$, the event horizon and\textcolor{white}{i}the Cauchy horizon coincide, indicating the extremal BH case.
 \end{justify}
\section{Geodesic, Innermost Stable and Innermost Bound Circular Orbits}
This section analyzes geodesics, ISCO, and IBCO radii, which are crucial for understanding the orbital dynamics around the Schwarzschild-MOG BH. These parameters provide the foundation for analyzing periodic orbits and their associated energy requirements in later sections.
\subsection{Geodesic Equations and Effective Potential}
We consider a test particle characterized by the trajectory $x^\mu \left(\tau \right)$, where $\tau$ serves as an appropriate affine parameter. To derive the equations of motion, we start with the Hamiltonian which is given by 
   \begin{align}
\label{eq6}\mathcal{H}=\frac{1}{2}g^{\mu\nu}p_\mu p_\nu=\frac{1}{2}\left(-\frac{p_t^2}{f}+fp_r^2+\frac{p_\theta^2}{r^2}+\frac{p_\phi^2}{r^2\sin^2\theta}\right),
\end{align}
where $ p_\mu=g_{\mu\nu}\Dot{x}^\nu$ represents the generalized conjugate momenta related to the 4-velocity $\Dot{x}^\mu$ by $p_t = g_{tt}\Dot{t}=-f\Dot{t}$, $p_r=g_{rr}\Dot{r}=\frac{\Dot{r}}{f}$, $p_\theta = g_{\theta\theta}\Dot{\theta}=r^2\Dot{\theta}$, and $p_\phi = g_{\phi\phi}\Dot{\phi}=r^2\sin^2\theta\Dot{\phi}$ (the dot represents the derivative with respect to $\tau$ along the geodesics). Since the Hamiltonian is a constant of motion and cyclic in $t$ and $\phi$,
the momenta along these directions are conserved, i.e.
$p_t=\frac{\partial\mathcal{H}}{\partial\Dot{t}}=-E$ and $p_\phi=\frac{\partial\mathcal{H}}{\partial\Dot{\phi}}=L$. Therefore, we have
\begin{equation}
p_t =-E,~ \implies\Dot{t}=\frac{E}{f},\qquad p_\phi =L,~\implies\Dot{\phi}=\frac{L}{r^2 \sin^2\theta},
\end{equation}
where $E$ and $L$ represents the total energy and total angular momentum of particle, which are constants of motion. Using Hamilton's equations ${\Dot{x}^\mu} = \frac{\partial \mathcal{H}}{\partial p_\mu}$ and ${\Dot{p}_\mu} = -\frac{\partial \mathcal{H}}{\partial x^\mu}$, we have
\begin{equation}
 \Dot{p}_t =\Dot{p}_\phi =0,\qquad \Dot{r} =f p_r,\qquad \Dot{\theta} =\frac{p_\theta}{r^2}.
 \end{equation}
The value of Hamiltonian is fixed by the normalization condition $\mathcal{H}=\frac{1}{2}g^{\mu\nu}p_{\mu} p_\nu=\frac{1}{2}\epsilon$, where $\epsilon=-1~(\epsilon=0)$ for massive (massless) particles. Therefore, we have
\begin{equation}
\Dot{p}_r =\frac{1}{2}\left[-p_r^2\left(f^\prime\right)+\frac{2\left(p_\theta^2\right)}{r^3}+\partial_r\left(\frac{R}{f}\right)-\frac{2\left(\Theta\right)}{r^3}\right],\qquad  \Dot{p}_\theta =\frac{1}{2r^2}\partial_\theta\left(\Theta\right),
\end{equation}
where $"\prime``$ represents the derivative with respect to $r$ and
\begin{equation}
R\left(r\right) =E^2-f\left(1+\frac{L^2}{r^2}+\frac{\mathcal{K}}{r^2}\right),\qquad
    \Theta\left(\theta\right) =\mathcal{K}-\frac{L^2\cos^2\theta}{\sin^2\theta}.
\end{equation}
Here, $\mathcal{K}$ is the separation constant. In the plane $\theta=\frac{\pi}{2}$, $\mathcal{K}=0$, and since we are considering the massive particles ($\epsilon=-1$), the effective potential is given by
\begin{align}
\Dot{r}^2=E^2-V_{\text{eff}}(r),
\end{align}
where
\begin{align}
V_{\text{eff}}=\left(1-\frac{2M\left(1+\alpha\right)}{r}+\frac{M^2\alpha\left(1+\alpha\right)}{r^2}\right)\left(\frac{L^2}{r^2}+1\right).
\end{align}
We see that as $r\rightarrow\infty$, $V_{\text{eff}}\rightarrow1$. The particles with $E>1$ can escape to infinity. $E=1$ shows the boundary between the bound and unbound orbits. Figure~\ref{fig 4} shows the plots of the effective potential for different values of $\alpha$ ($\alpha=0$ represents the Schwarzschild BH case).
 \subsection{Innermost Stable Circular Orbits}
 The stability of a circular orbit is determined by checking the signature of $\frac{d^2 }{dr^2}\left(V_{\text{eff}}\right)$ at the orbital radius $r_0$. For a stable orbit $\frac{d^2}{dr^2} \left(V_{\text{eff}}\right)|_{r=r_0}>0$ and unstable one has $\frac{d^2 }{dr^2}\left(V_{\text{eff}}\right)|_{r=r_0}<0$. ISCO represents the closest orbit around a massive object where a test particle can orbit stably. For ISCO, the conditions are \cite{bambi2017black}
\begin{align}
\frac{d}{dr}\left(V_{\text{eff}}\right)=0,~\frac{d^2}{dr^2}\left(V_{\text{eff}}\right)=0,~V_{\text{eff}}=E^2.
\end{align}
To determine the ISCO, which represents the innermost and stable bounded orbit, one must have $E<1$. For the Schwarzschild-MOG BH, these conditions yield
\begin{equation}
L =r \sqrt{\frac{M\left(1+\alpha\right)\left(r-M\alpha\right)}{r^2-3Mr-3M\alpha r+2M^2\alpha+2M^2\alpha^2}},\quad  E =\frac{r^2-{2Mr\left(1+\alpha\right)}+{M^2\alpha\left(1+\alpha\right)}}{r\sqrt{r^2-3Mr\left(1+\alpha\right)+2M^2\alpha\left(1+\alpha\right)}},
\end{equation}
with the radius of ISCO \cite{boboqambarova2023particle}
\begin{align}
r_{\text{ISCO}}=M\left(2\left(1+\alpha\right)+{Y}+\frac{\left(4+\alpha\right)\left(1+\alpha\right)}{Y}\right),\end{align}where\begin{align}{Y}=\sqrt[3]{\left(1+\alpha\right)\left(8+7\alpha+\alpha^2+\alpha\sqrt{\alpha+5}\right)}.
\end{align}
This result shows that increasing $\alpha$ increases the value of $r_{\text{ISCO}}$. When the MOG parameter $\alpha$ is absent ($\alpha=0)$, all these quantities reduce to those of the Schwarzschild BH, with $r_{\text{ISCO}}=6M$. When the denominator of the energy is zero, we obtain the radius of the circular orbit, given as \cite{hu2022observational}
\begin{align}
\label{eq17}r_\text{ph} =\frac{M}{2}\left(3+3\alpha+\sqrt{9+10\alpha+\alpha^2}\right).
        \end{align}
        It corresponds to the radius of the region around a BH where light (massless particles) orbits in a circular path.
        From Eq.~\eqref{eq17} we observe that, as $\alpha$ increases, $r_\text{ph}$ also increases.
\subsection{Innermost Bound Circular Orbits}
For IBCO, the conditions for particle motion are \cite{bambi2017black}
    \begin{align}
    \frac{d}{dr}\left(V_{\text{eff}}\right)=0,~V_{\text{eff}}=E^2=1.
    \end{align}
Solving these conditions gives the radius of the IBCO, given as
\begin{align}
    r_{\text{IBCO}} =\frac{M}{3}\left[4\left(1+\alpha\right)+\frac{4\sqrt[3]{2}\left(1+\alpha\right)\left(4+\alpha\right)}{Z}+\frac{Z}{\sqrt[3]{2}}\right],
\end{align}
where
\begin{align}
Z=(128+240\alpha+123\alpha^2+11\alpha^3+3\sqrt{3}\sqrt{-32\alpha^3-69\alpha^4-42\alpha^5-5\alpha^6})^\frac{1}{3}.
\end{align}
\begin{figure}[H]
 \centering
{{\includegraphics[height=10.5cm,width=13.7cm]{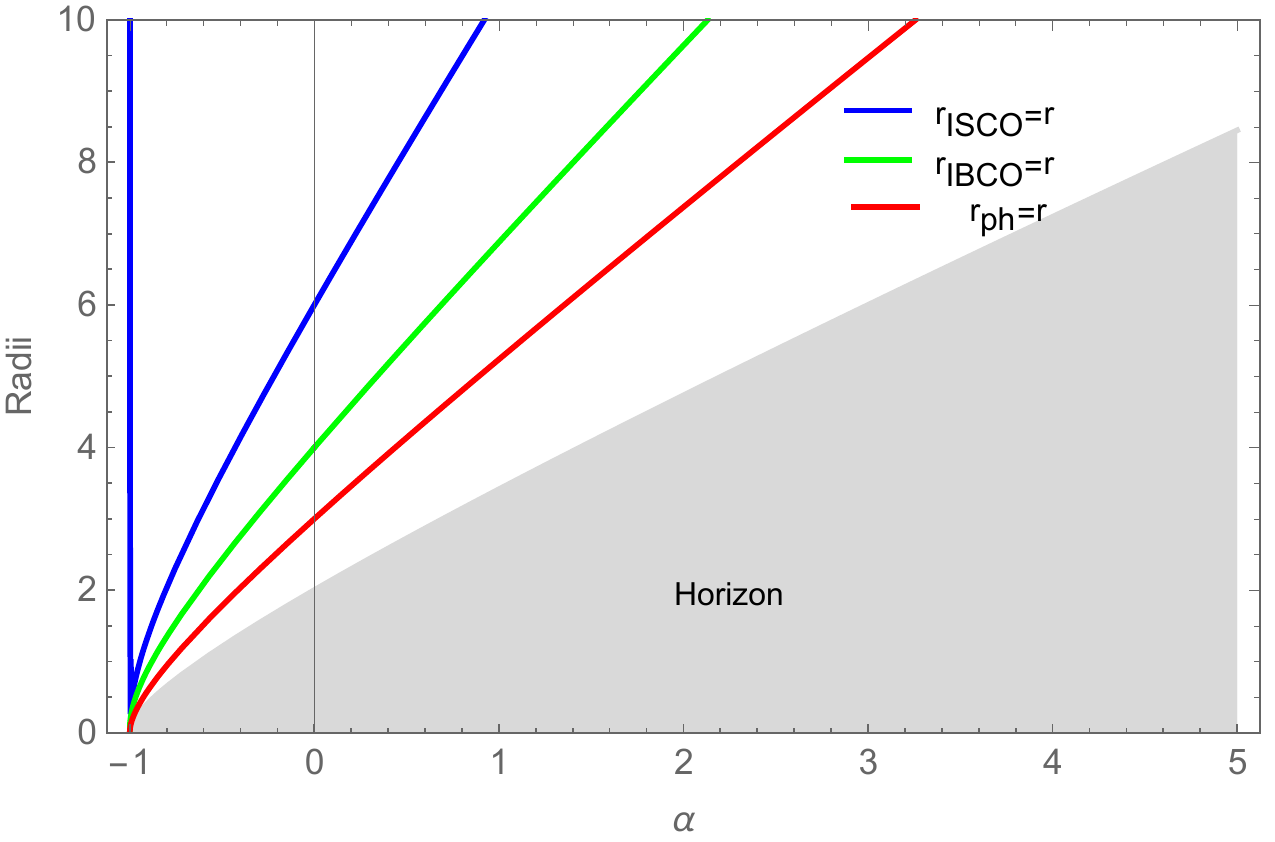}}}
\caption{The characteristic radii, namely, the horizon, photon sphere,
IBCO and ISCO position as the functions of $\alpha$ parameter.}
\label{fig 2}
 \end{figure}\begin{justify} 
 It represents the innermost and unstable bounded orbit with $E = 1$. When $\alpha$ is absent $(\alpha = 0)$,
$r_\text{{IBCO}} = 4M$, which is the radius of the IBCO in Schwarzschild spacetime. \\\indent In IBCO, one massive particle has the same energy as the particle's energy at infinity. It means there is no energy dissipation when the massive particle moves from infinity to the IBCO. Fig.~\ref{fig 2} shows the dependence of radii from the $\alpha$ parameter. We see that the characteristic radii, namely horizon, photon sphere, IBCO, and ISCO radii for test particles get large in the presence of MOG parameter $\alpha$. In the limit $\alpha=-1$, the radii of the horizon, photon sphere, and IBCO become zero but the ISCO becomes infinite. To support a bound orbit, the energy cannot be too high, as the particle would escape to infinity, nor can it be too low, as it would fall into the BH. The energy depends on the angular momentum $L$ and is restricted by $E_{\text{ISCO}}^2\leq E^2\leq E_{\text{IBCO}}^2=1$ for a given particle. To reduce the number of free parameters, we consider a new angular momentum given as
\begin{align}\label{av}
L_{\text{av}}=\frac{L_{\text{ISCO}}+L_{\text{IBCO}}}{2}.
\end{align}
\begin{figure}[H]
 \centering
{{\includegraphics[height=7.5cm,width=7.5cm]{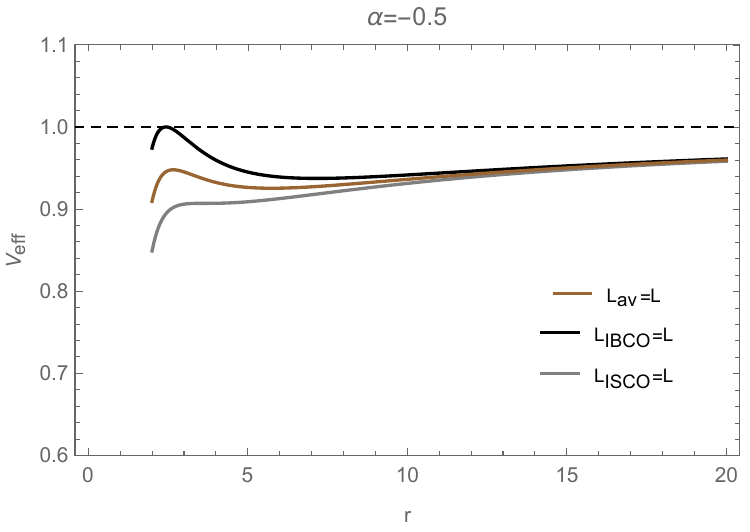}}}
{{\includegraphics[height=7.5cm,width=7.5cm]{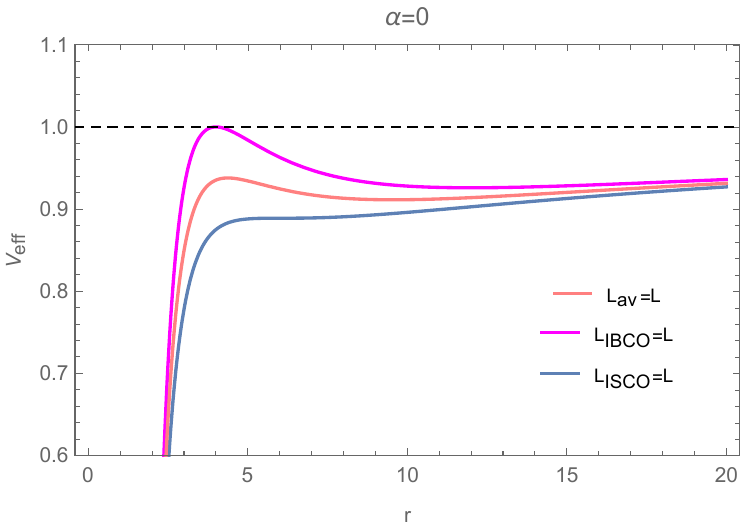}}}\end{figure}\begin{figure}[H]
 \centering
{{\includegraphics[height=7.5cm,width=7.5cm]{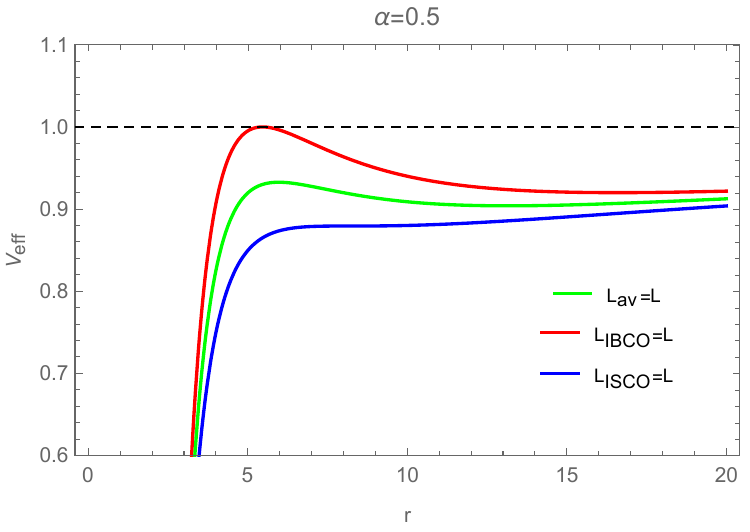}}}
{{\includegraphics[height=7.5cm,width=7.5cm]{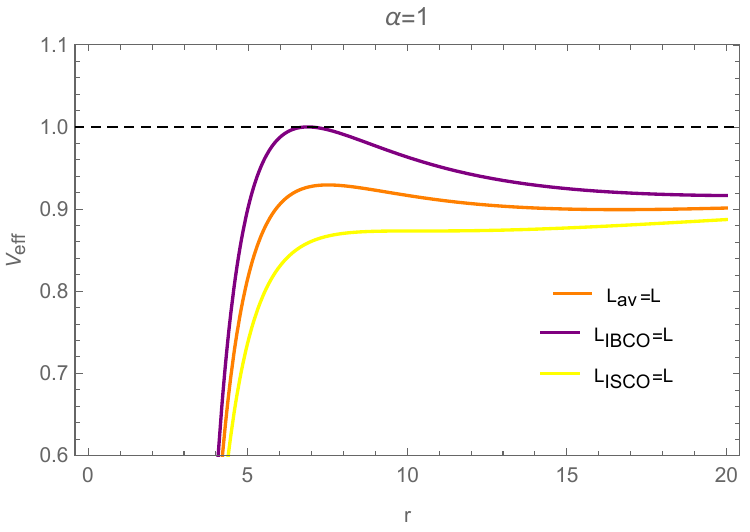}}}
\caption{Behavior of effective potential with angular momentum for $M=1$, and various $\alpha$.}
\label{fig 3}
 \end{figure}\end{justify}
 \begin{justify}
For each $\alpha$, we can uniquely determine $L_{\text{ISCO}}$ and $L_{\text{{IBCO}}}$, where the angular momentum of the particles satisfy $L_{\text{ISCO}}<L<L_{\text{IBCO}}$ \cite{levin2008periodic}. Fig.~\ref{fig 3} shows the behavior of effective potential for different angular momentum. For $L_{\text{IBCO}}$, the peak of the potential barrier coincides with $V_{\text{eff}}=1$ for every $\alpha$. We observe that for $\alpha\geq0$, the effective potential exhibits a deep potential well,  whereas for $-1<\alpha<0$, a potential well still forms but becomes shallower and smaller in size. This deep potential well allows for a rich variety of periodic orbits. Therefore, to obtain periodic orbits, we consider $\alpha$ in the range $(-1,\infty)$. \end{justify}
\section{Periodic Orbits}
In this section, we will explore the periodic orbits of test particles around the Schwarzschild-MOG BH, following the taxonomy introduced by Levin \cite{levin2008periodic}. Any bound orbit can be characterized by its oscillations in the radial coordinate $r$ and the angular coordinate $\phi$. A periodic orbit occurs when the ratio of these orbital frequencies forms a rational number, i.e., $q=\frac{\omega_\phi}{\omega_r}$, such that the trajectory closes and the particle returns to its starting point within a finite time, repeating its motion. The positive rational number $q$ can be decomposed into three integers $\left( z,w,v \right)$, which can be written as
\begin{align}
   \label{eq33} q=w+\frac{v}{z}=\frac{\Delta \phi}{2 \pi}-1,
    \end{align}
    where
    \begin{align}
        \Delta\phi=\oint d\phi,
    \end{align}
represents the equatorial accumulated angle for periodic orbits between two successive apastra \cite{tu2023periodic}. The triplet of integers can be geometrically interpreted as describing the structure of the trajectory, where $z$ denotes the ``zoom" number, $w$ represents the number of ``whirls", and $v$ is the ``vertex" number. For periodic orbits, the evolution of $\Delta\phi$ must be an integer multiple of $2\pi$; hence, we have
\begin{align}
      \Delta\phi &=2\int_{\phi_1}^{\phi_2} d\phi,\nonumber\\
&=2\int_{r_-}^{r_+}\frac{\Dot{\phi}}{\Dot{r}}dr,\nonumber\\
\label{eq35}&=2\int_{r_-}^{r_+}\frac{L}{r^2\sqrt{E^2-f\left(1+\frac{L^2}{r^2}\right)}}dr,
\end{align}
where $\phi_1$ and $\phi_2$ denote periastron and apastron. Here, $r_+$ and $r_-$ are two turning points between the ISCO and IBCO, which can be determined by the expression $\Dot{r}^2=0$, where $r_\pm$ are the roots of $E^2=V_{\text{eff}}$. Now, using Eqs.~\eqref{eq33} and~\eqref{eq35}, the rational number $q$ can be written as
\begin{align}
    \label{eq36}q =\frac{1}{\pi}\int_{r_-}^{r_+}\frac{L}{r^2\sqrt{E^2-f\left(1+\frac{L^2}{r^2}\right)}}dr-1.
\end{align}
According to Eq.~\eqref{eq36}, the rational number $q$ depends on $E$, $L$ and $\alpha$ when $M=1$. If we further consider $L=L_{\text{av}}$, where the value of $L_{\text{av}}$ is fixed by $\alpha$ and $M$, then $q$ depends only on $E$. We can calculate the value of $q$ numerically for different values of $\alpha$.
 \begin{figure}[H]
 \centering
{{\includegraphics[height=7cm,width=7.5cm]{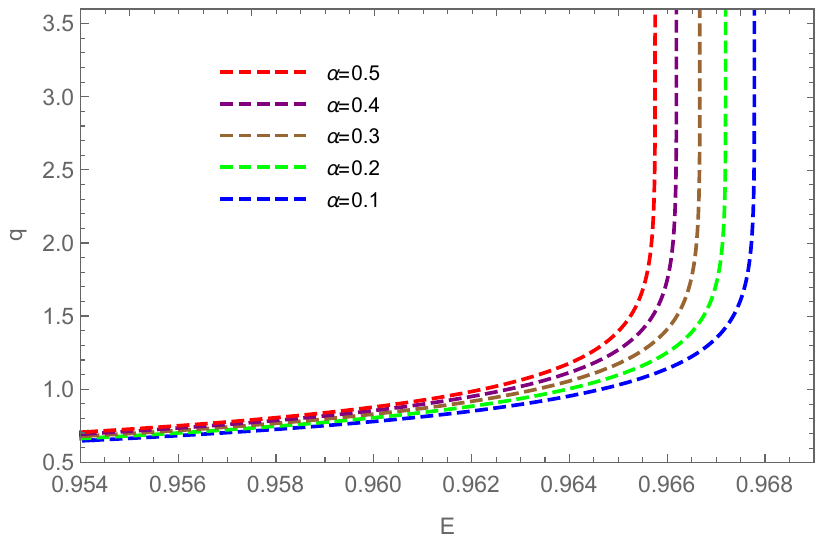}}}
{{\includegraphics[height=7cm,width=7.5cm]{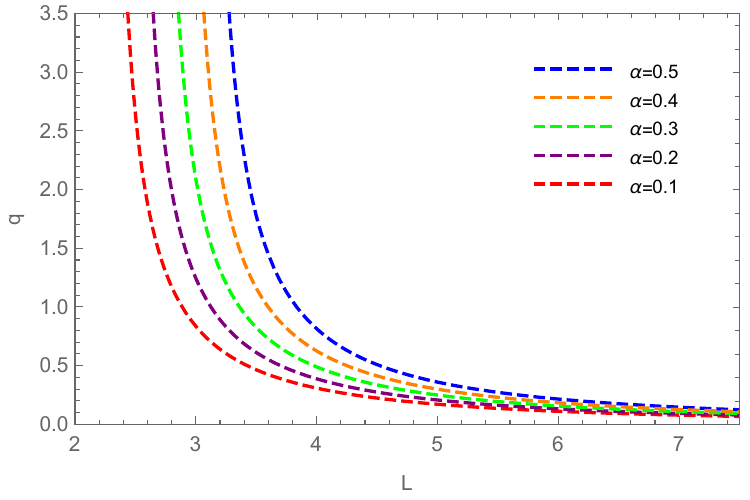}}}
\caption{Variation of $q$ as a function of energy and angular momentum. For the left plot, $\alpha=0.1,0.2,0.3,0.4,0.5$, with their corresponding $L_{\text{av}}$. The right plot shows the same $\alpha$ values, with the energy kept fixed for the $(1,1,0)$ orbit.}\label{fig 4}
\end{figure}
\begin{justify}Fig.~\ref{fig 4} shows the variation of $L$ and $E$ with $q$. In the left plot, by adjusting the value of $E$ for a given angular momentum, we obtain a series of periodic orbits. For each $\alpha$, we see that as $E$ increases to a certain value, $\Delta\phi$ and $q$ goes to infinity. This means that $q$ increases with the particle's energy. If we instead fix energy and vary $L$,  the opposite behavior is observed in the right plot. Using Eq.~\eqref{eq36}, we construct Tables~\ref{tab:1} and~\ref{tab:2} for different values of $\alpha$. By analyzing both tables, we see that for $\alpha>0$, the energy values of periodic orbits around Schwarzschild-MOG BH are lower than the energy values of the Schwarzschild BH. However, for $-1<\alpha<0$, the energy of periodic orbits around the Schwarzschild-MOG BH is higher than that for the Schwarzschild BH. \end{justify}
     \begin{table}[H]
     \begin{center}
     \caption{Energy values of $(z=1,2,3,w=1,v=1)$ periodic orbits around Schwarzschild-MOG BH for $\alpha=-0.5,-0.4,-0.3,-0.2,-0.1$ with their corresponding angular momentum.}
     \label{tab:1}
    \begin{tabular}{cccccc}
    \toprule
      $\alpha$ & $L_{\text{av}}$ &$E_{(1,1,0)}$  &$E_{(2,1,1)}$  & $E_{(3,1,1)}$ & $E_{(4,1,1)}$\\
      \midrule
        -0.5 & 2.07745 &0.971211  & 0.973257 & 0.972964 &0.972723  \\
         -0.4 & 2.41579 & 0.969673  & 0.971845  & 0.971532 & 0.971276  \\ 
        -0.3 &2.74946  & 0.968378  & 0.970667  & 0.970335    & 0.970064  \\
         -0.2 & 3.079585 & 0.967262  &0.969662  & 0.969313 & 0.969028\\
         -0.1 & 3.40693 &0.967262  & 0.969662 & 0.969313 &0.969028 \\
          \bottomrule
    \end{tabular}
    \end{center}
    \end{table}
\begin{table}[H]\begin{center}
    \caption{Energy values of $(z=1,2,3,w=1,v=1)$ periodic orbits around Schwarzschild-MOG BH for $\alpha=0,0.1,0.2,0.3,0.4,0.5$ with their corresponding angular momentum. Similar results are obtained upto $\alpha=1$. }
     \label{tab:2}
    \begin{tabular}{cccccc}
    \toprule
      $\alpha$ & $L_{\text{av}}$ &$E_{(1,1,0)}$  &$E_{(2,1,1)}$  & $E_{(3,1,1)}$ & $E_{(4,1,1)}$\\
      \midrule
        0 & 3.732055 &0.965426  & 0.968027 & 0.967645 &0.967334 \\
         0.1& 4.055345 & 0.964655  &  0.967349 & 0.966951  
 & 0.966629 \\ 
        0.2 & 4.37711 & 0.963962  & 0.966742 & 0.96633   & 0.965996 \\
         0.3& 4.697595 & 0.963333 &0.966196  & 0.96577  & 0.965426\\
         0.4& 5.016975 & 0.962759 & 0.9657  & 0.96526 & 0.964906 \\
         0.5& 5.3354 & 0.962232  & 0.965247 &0.964795  &0.964432 \\
         \bottomrule
    \end{tabular}
    \end{center}
    \end{table}
    \begin{justify}
     \begin{figure}[H]
 \centering
{{\includegraphics[height=4.5cm,width=5cm]{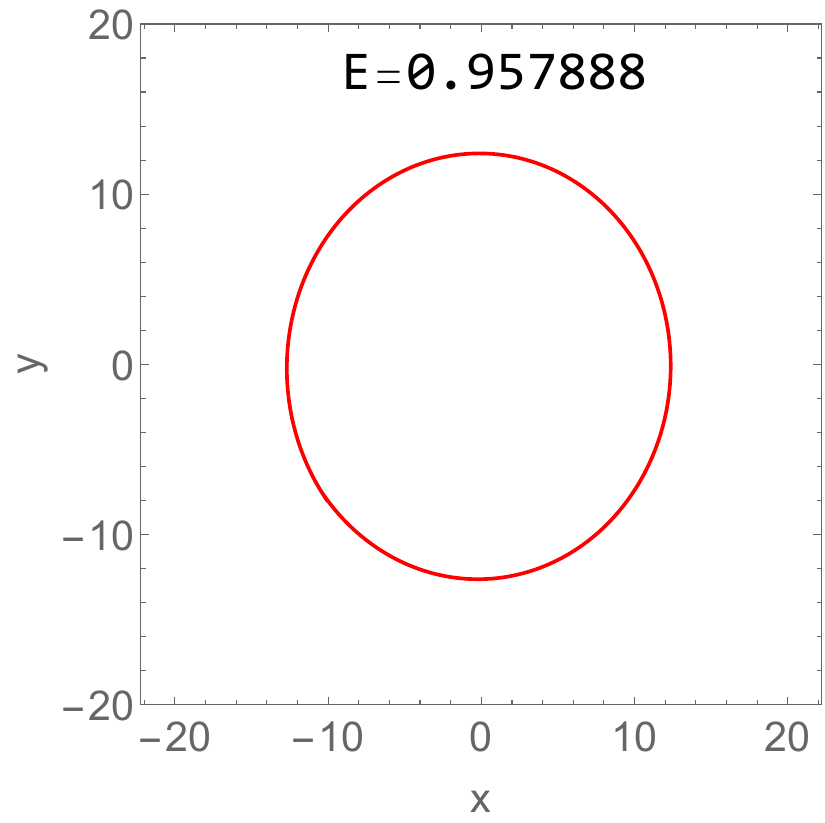}}}
{{\includegraphics[height=4.5cm,width=5cm]{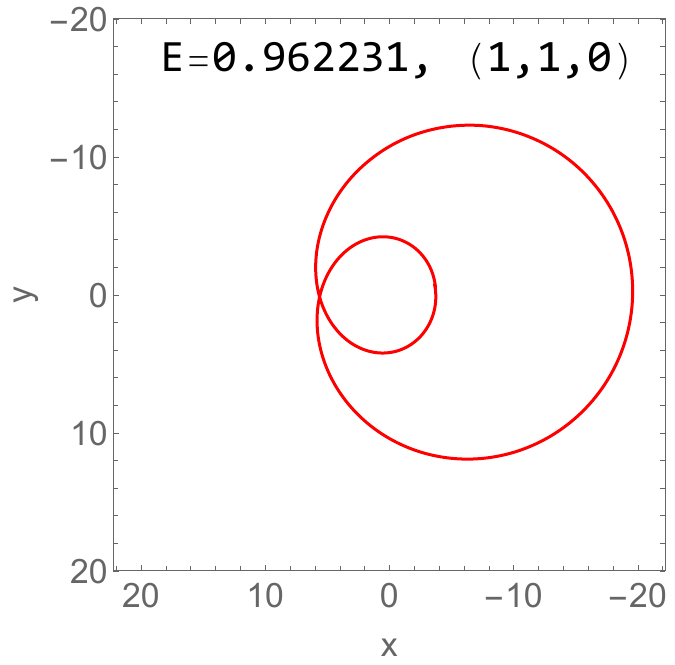}}}
{{\includegraphics[height=4.5cm,width=5cm]{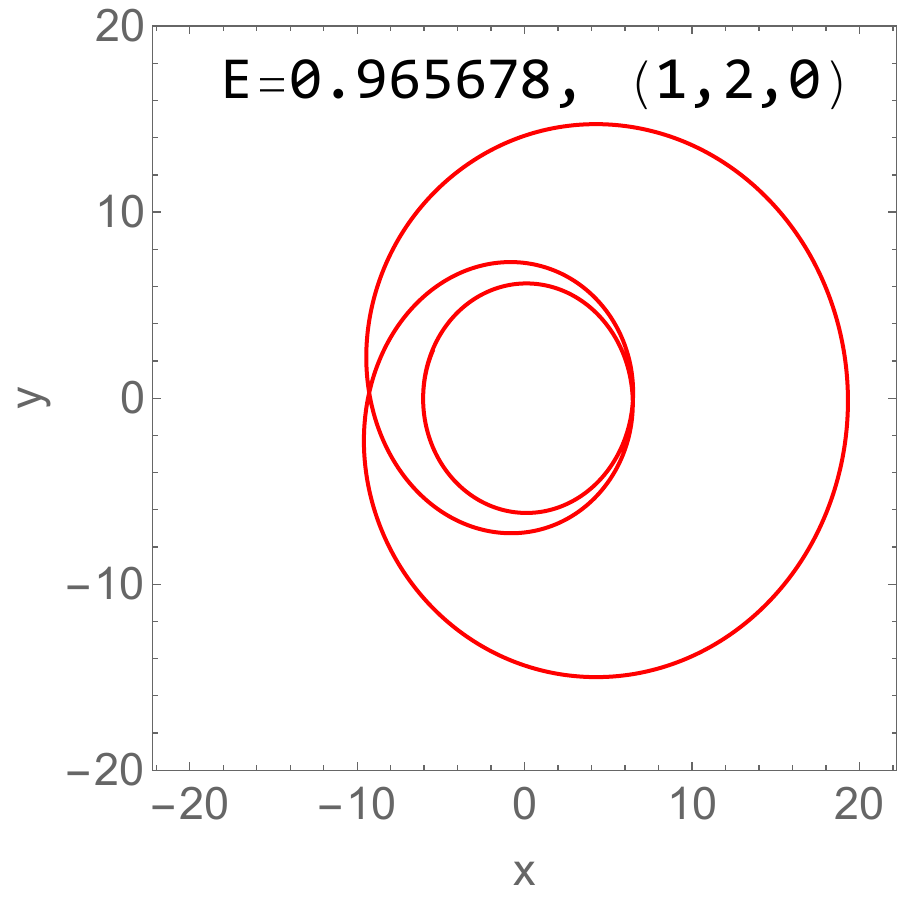}}}
 {{\includegraphics[height=4.5cm,width=5cm]{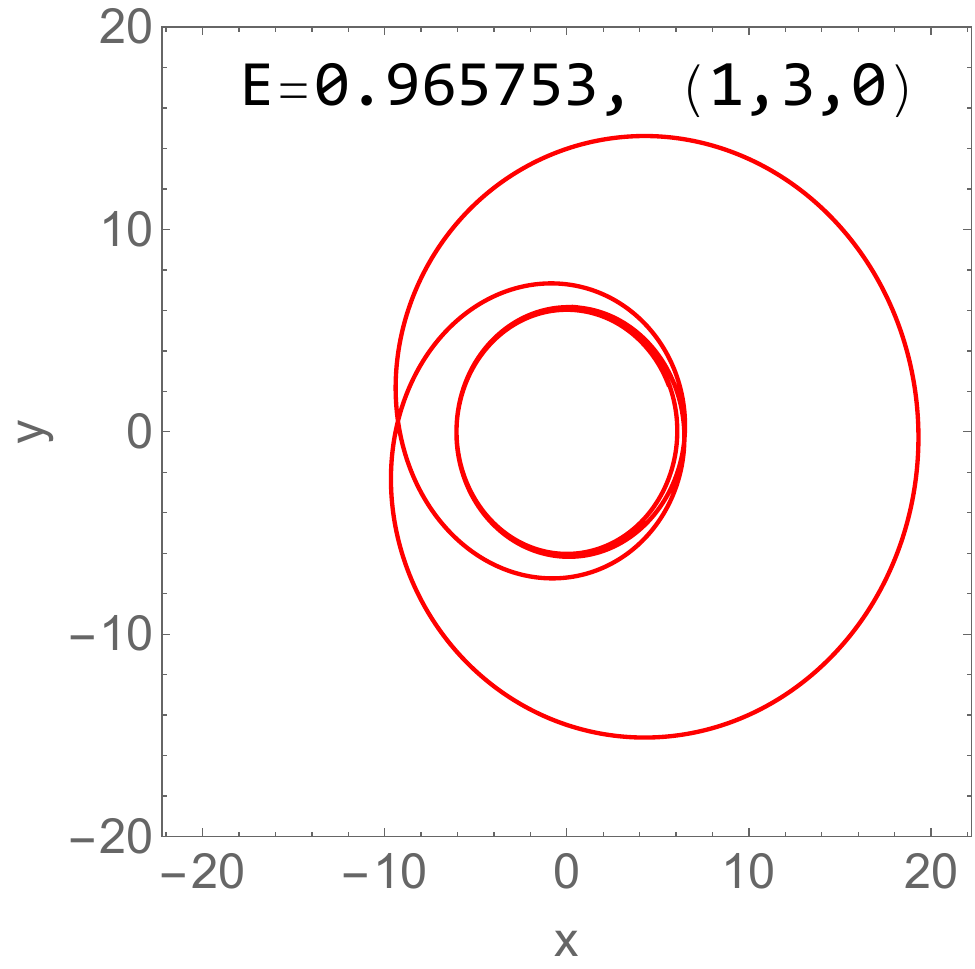}}}
{{\includegraphics[height=4.5cm,width=5cm]{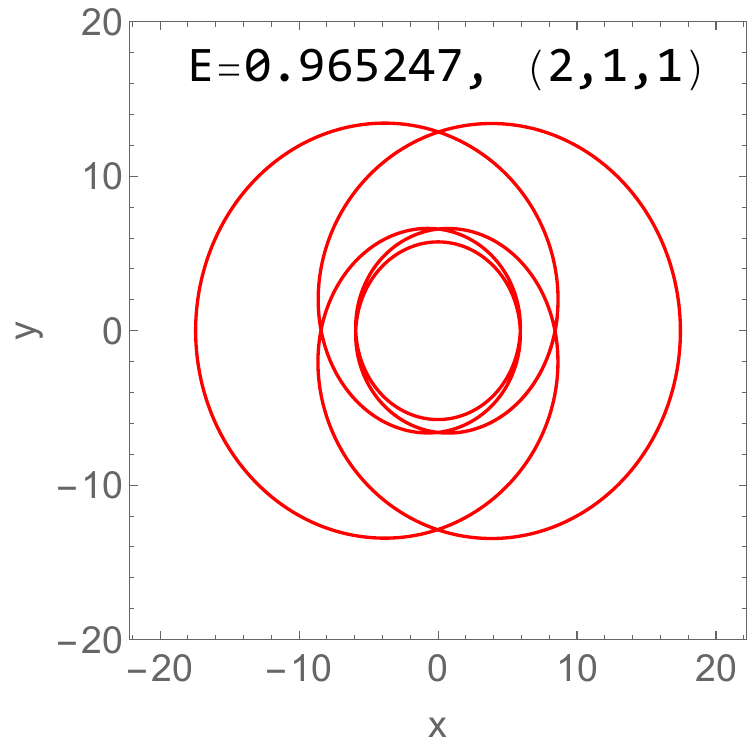}}}
{{\includegraphics[height=4.5cm,width=5cm]{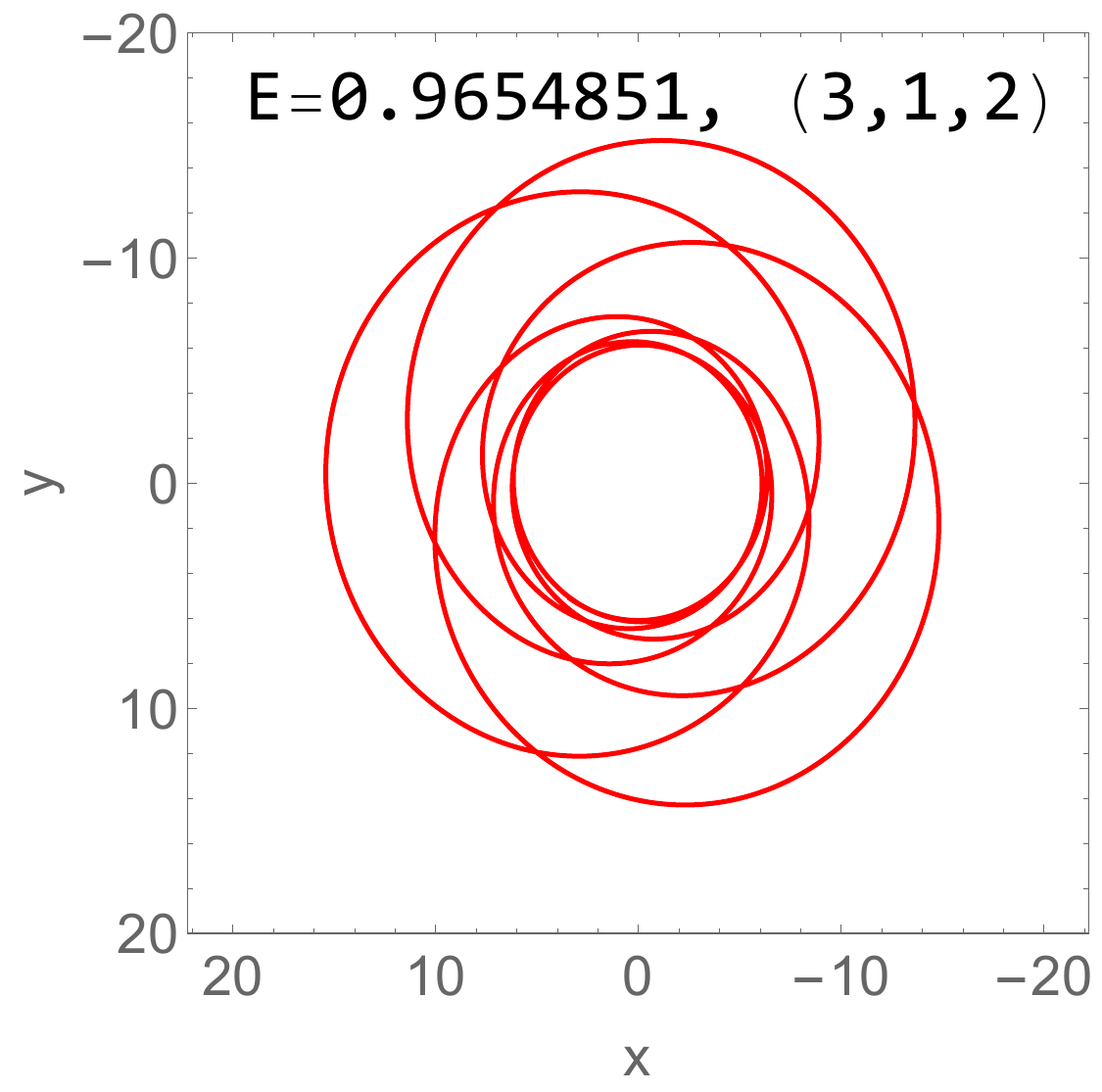}}}
{{\includegraphics[height=4.5cm,width=5cm]{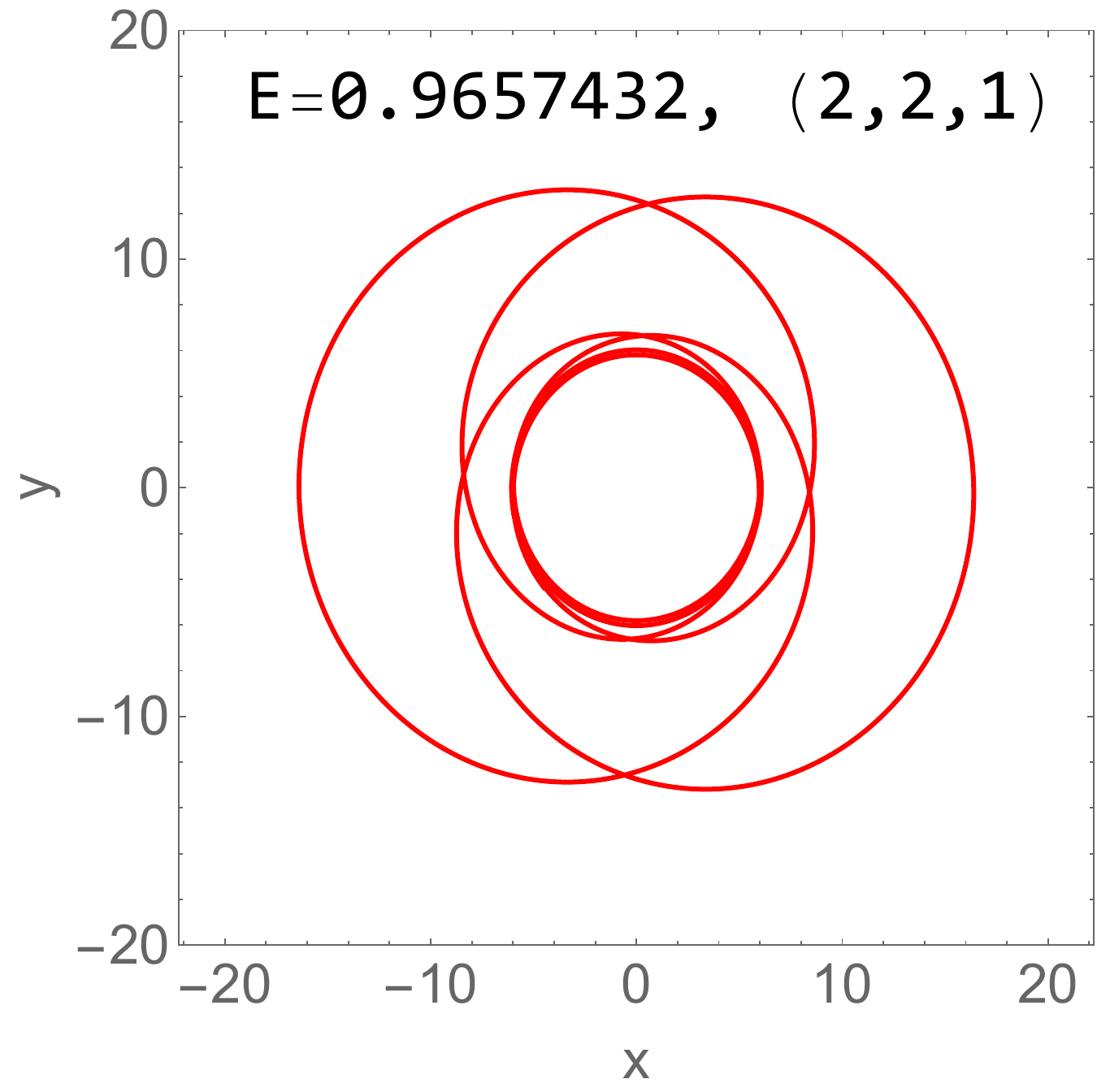}}}
{{\includegraphics[height=4.5cm,width=5cm]{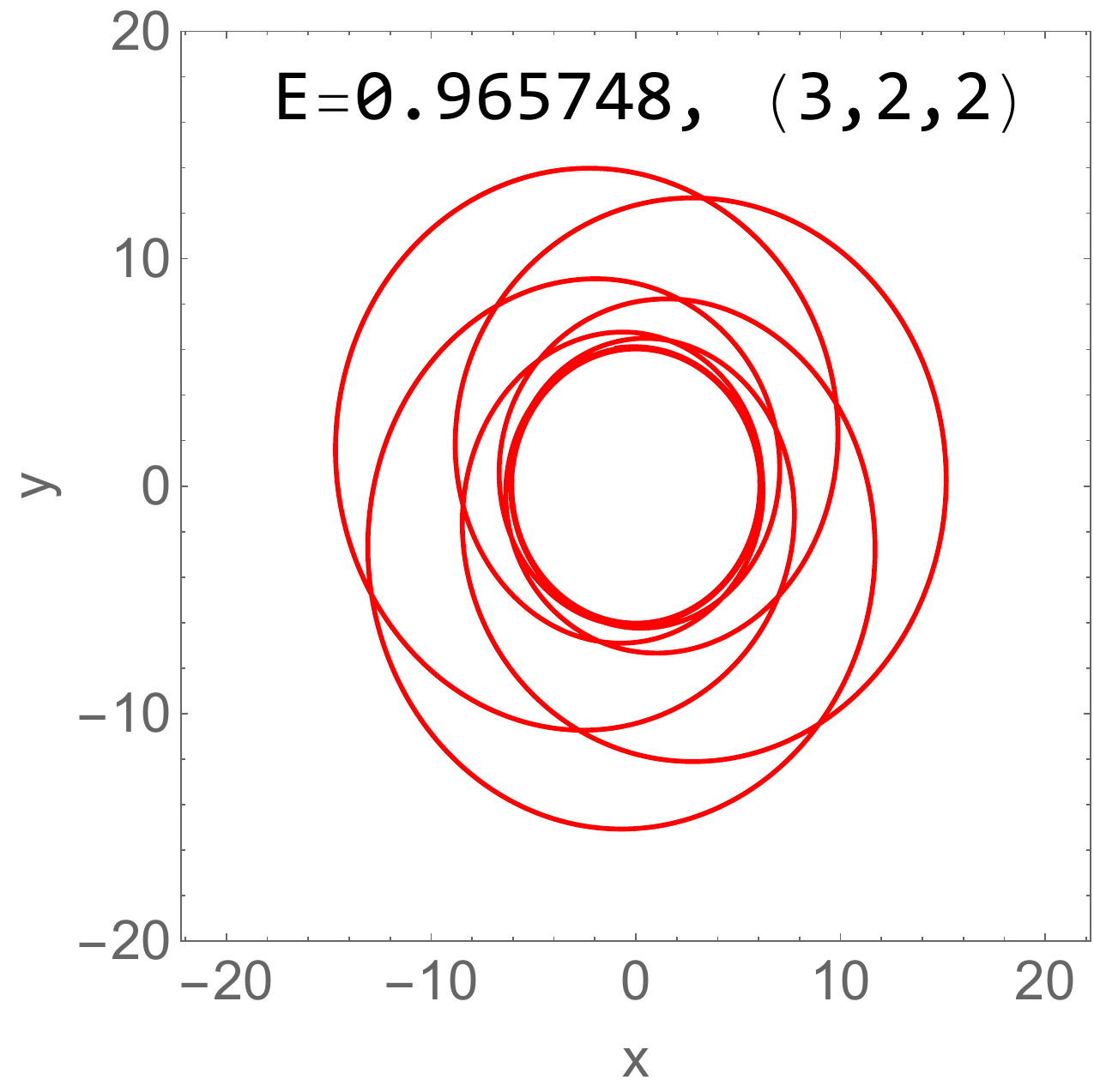}}}
{{\includegraphics[height=4.5cm,width=5cm]{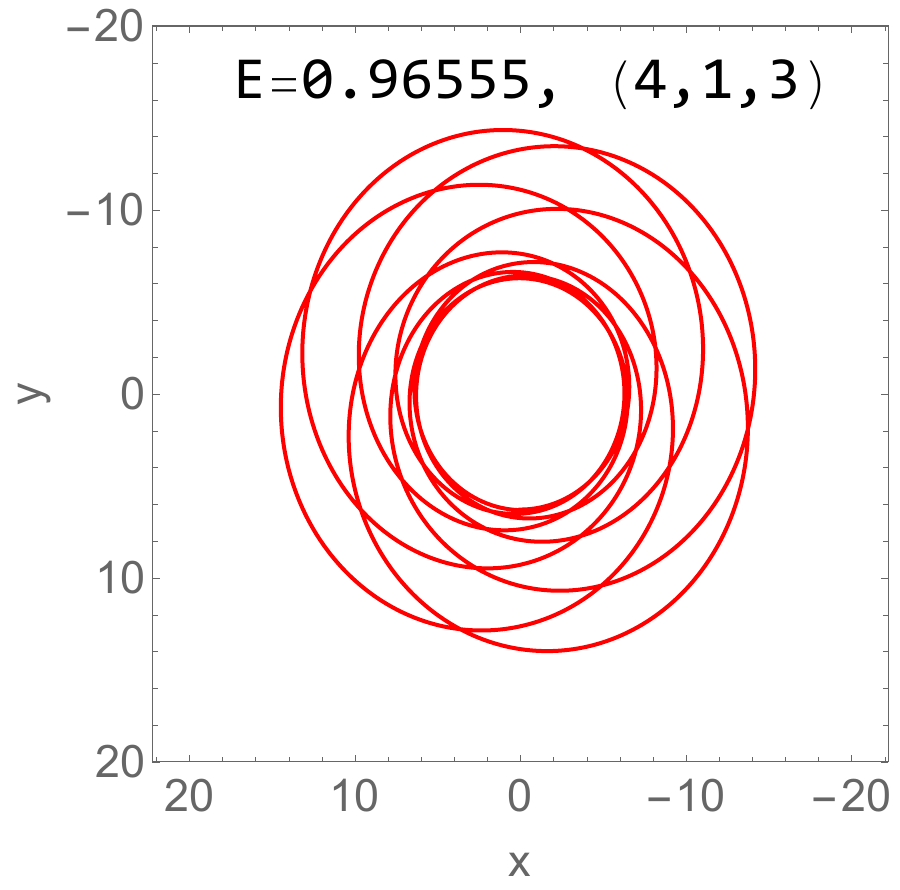}}}
\caption{Zoom-whirl periodic orbits characterized by $(z,w,v)$ with $z=1,2,3,4$, $w=1,2,3$, $v=1,2,3$, and $\alpha=0.5$. Here, $x=r\cos\phi$ and $y=r\sin\phi$.}\label{fig 5}
 \end{figure}\end{justify}
 \begin{justify}
 Fig.~\ref{fig 5} shows the periodic orbits of massive particles between the ISCO and IBCO around the\\ Schwarzschild-MOG BH for $
\alpha=0.5$. Each periodic orbit with a triplet of integers corresponds to different energy values. We can also analyze periodic orbits for other values of $\alpha$ with a triplet of integers. Fig.~\ref{fig 6} shows $(1,1,0)$, $(1,2,0)$, $(2,1,1)$, and $(3,1,1)$ orbits for $\alpha=1$.
     \begin{figure}[H]
 \centering
{{\includegraphics[height=7.2cm,width=7.5cm]{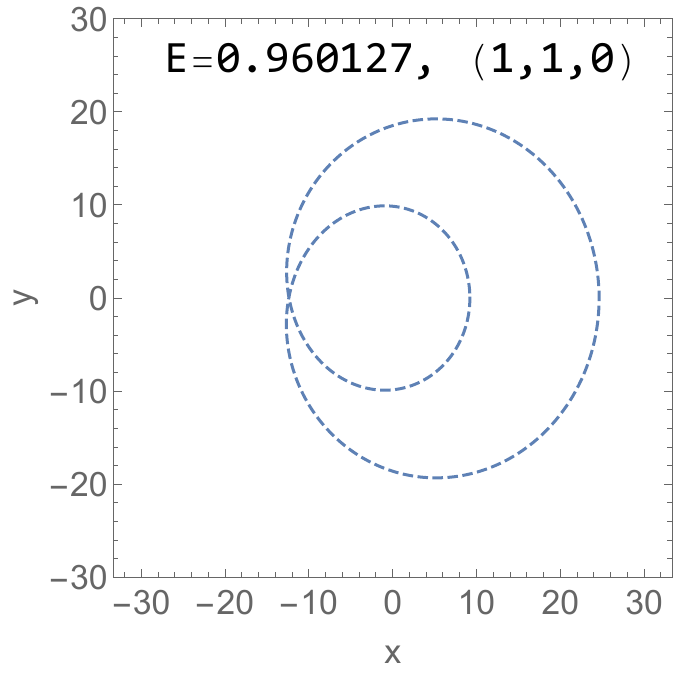}}}
{{\includegraphics[height=7.2cm,width=7.5cm]{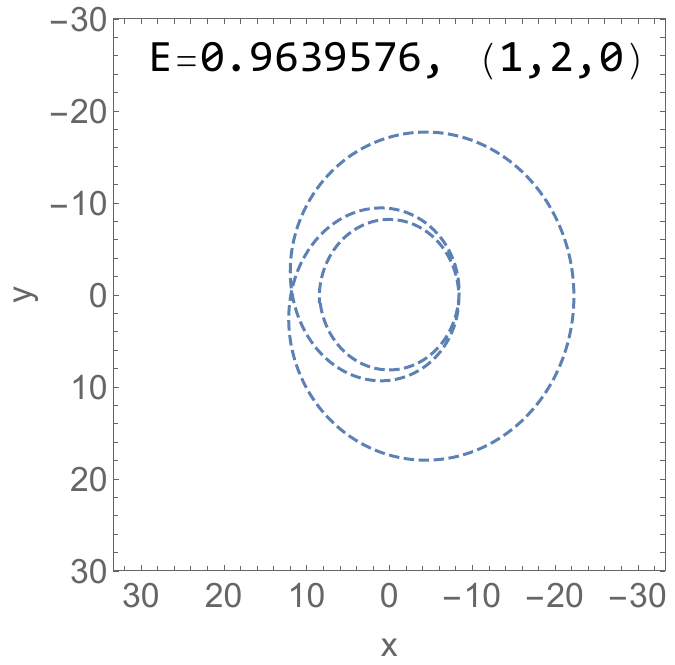}}}
{{\includegraphics[height=7.2cm,width=7.5cm]{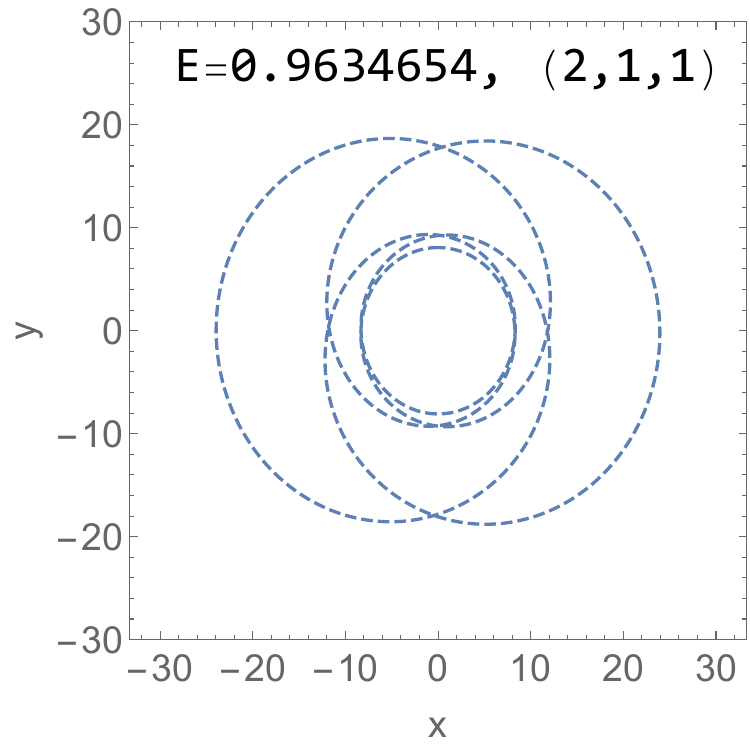}}}
{{\includegraphics[height=7.2cm,width=7.5cm]{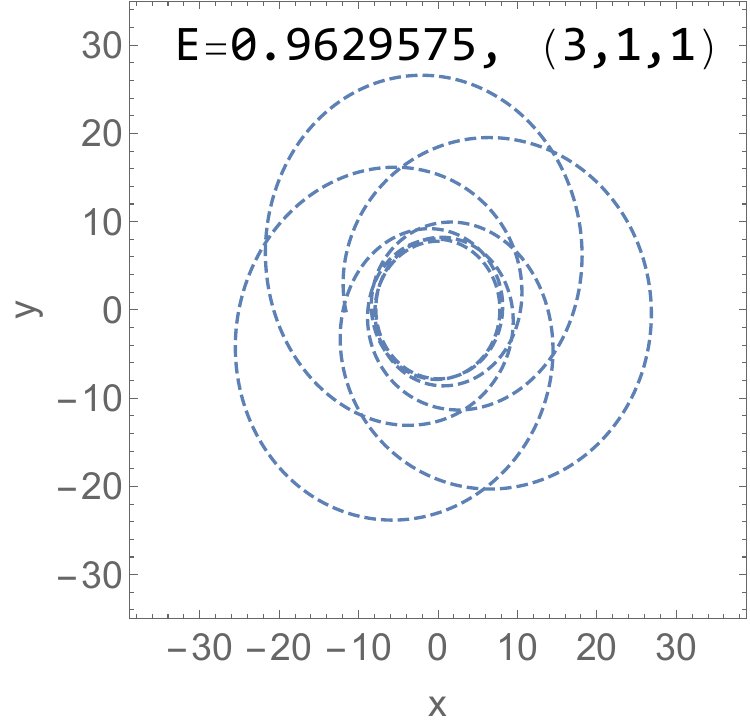}}}
\caption{Zoom-whirl periodic orbits characterized by $z=1,2,3,~w=1,2,$ and $v=1$ for $\alpha=1$.}\label{fig 6}
\end{figure}
\begin{justify}After a thorough analysis, we conclude that periodic orbits in Schwarzschild-MOG spacetime exhibit qualitative similarities to those in Schwarzschild and Kerr spacetimes. However, we find that achieving periodic orbits in Schwarzschild spacetime requires higher energy compared to Schwarzschild-MOG spacetime.\end{justify}
\end{justify}
\section{Precession of Periodic Orbits}
Generic orbits are not periodic and do not correspond to a rational $q$. However, any generic orbit can be approximated by a nearby periodic orbit, just as an irrational number can be approximated by a nearby rational number \cite{healy2009zoom}. The precession of periodic orbits refers to the gradual shift in the orbit's orientation, caused by small perturbations in the system. Considering a small perturbation in periodic orbits, where the irrational numbers can be defined as \cite{wang2022periodic}
\begin{align}
    q=w+\frac{v}{z}\pm\delta,
\end{align}
where $0<\delta\ll1$. We employed the prior analysis to visualize precessing orbits near periodic orbits with different energy values. Fig.~\ref{fig 7} illustrates several precessing orbits with different energy values.
 \begin{figure}[H]\label{fig 3.81}
 \centering
{{\includegraphics[height=7cm,width=7.3cm]{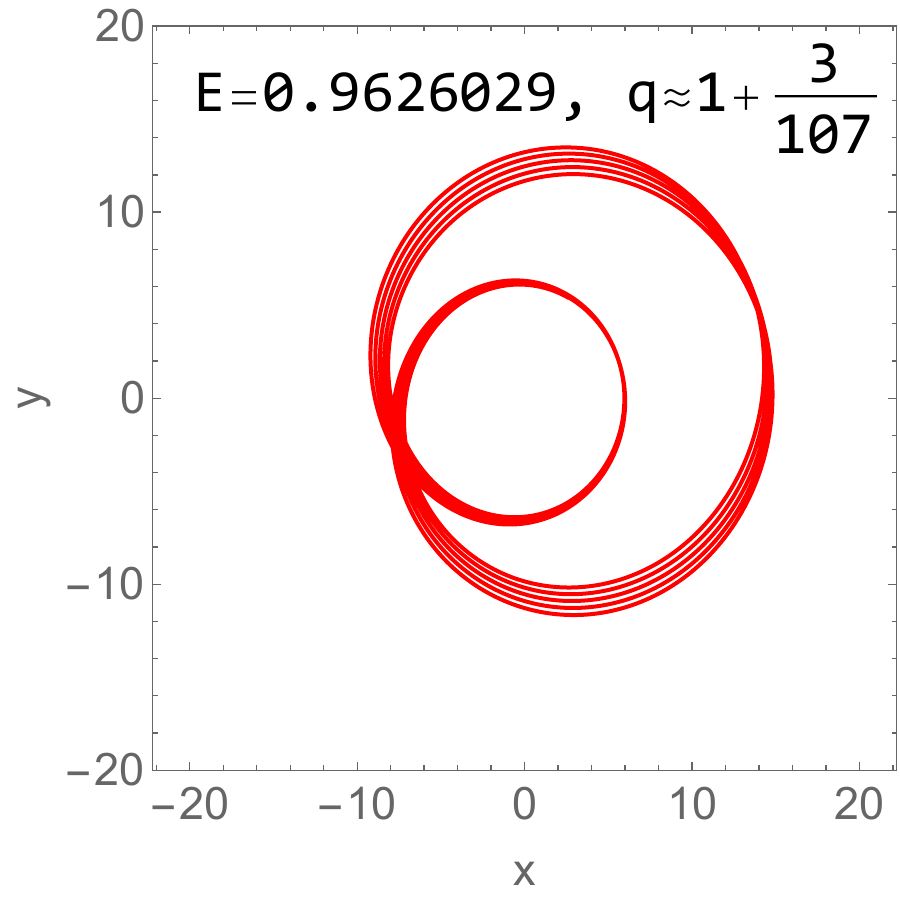}}}
{{\includegraphics[height=7cm,width=7.3cm]{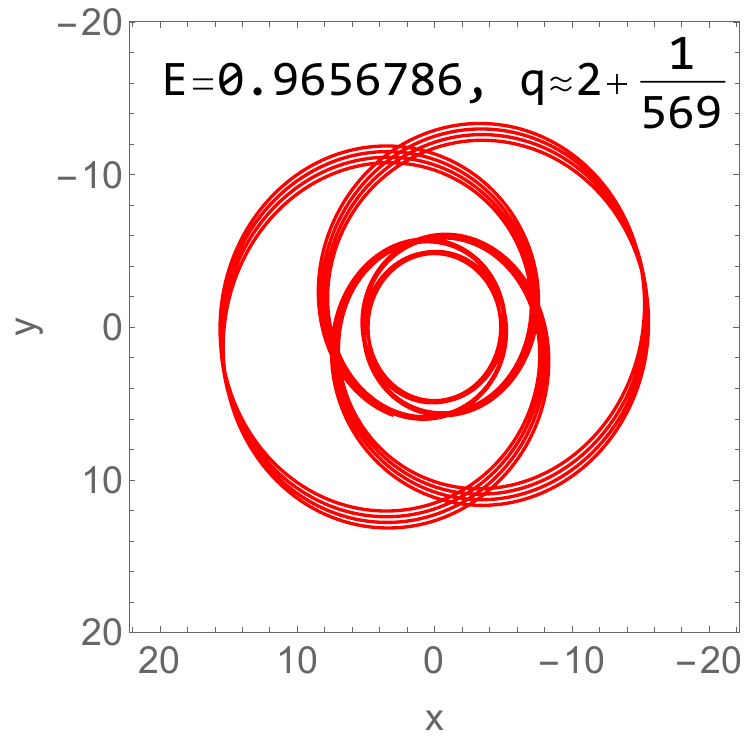}}}
{{\includegraphics[height=7cm,width=7.3cm]{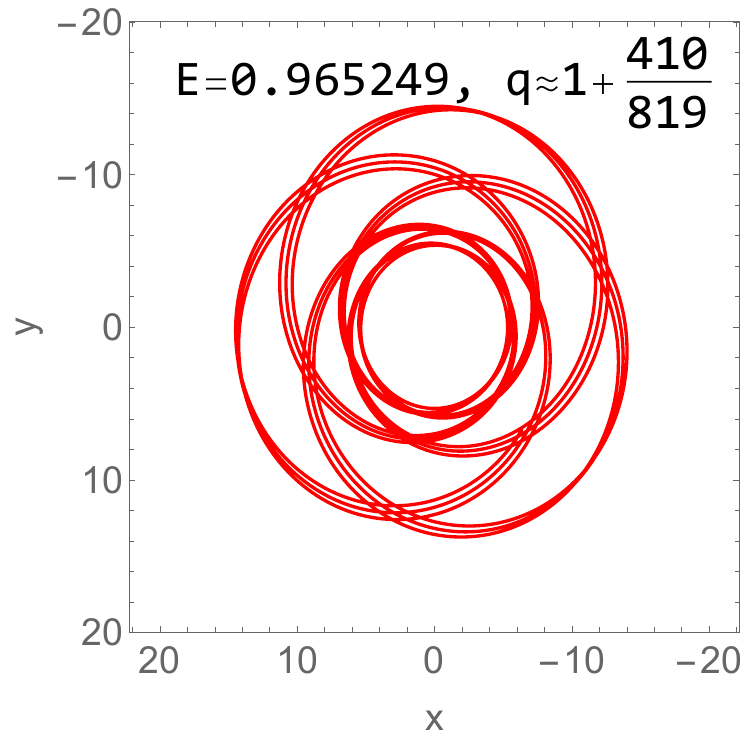}}}
{{\includegraphics[height=7cm,width=7.3cm]{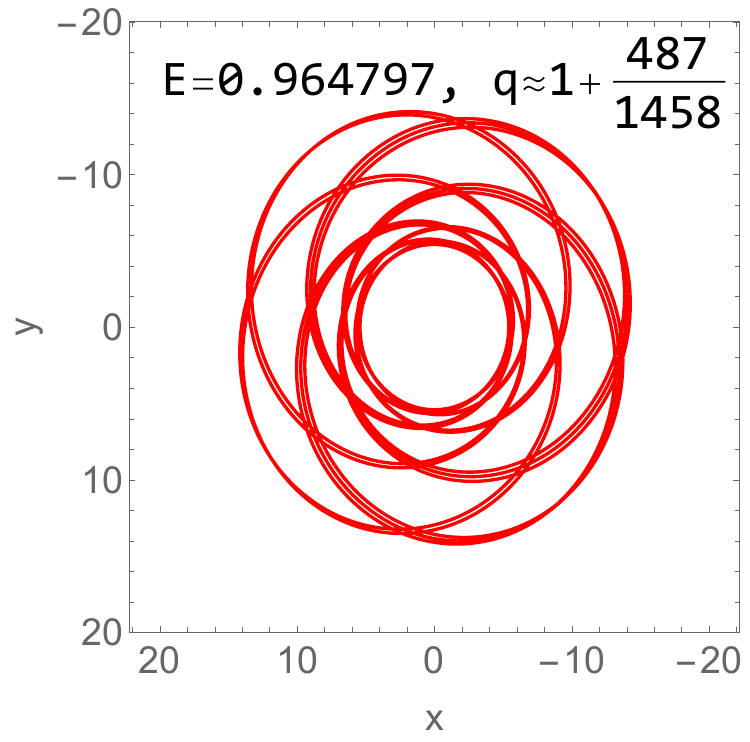}}}
\caption{Precessing orbits with varying energy values near periodic orbits for $\alpha=0.5$ and $M=1$, showing the gradual shift in orientation due to small perturbations.}\label{fig 7}
\end{figure}
\section{Gravitational Wave Radiation
 from Periodic Orbits}
 In this section, we explore the gravitational radiation emitted by the periodic orbits of a test particle orbiting a supermassive BH within the framework of MOG. To achieve this, we consider an EMRI system, where the smaller object’s mass is significantly smaller than that of the supermassive BH. This allows us to treat the smaller object as a perturbation to the spacetime of the Schwarzschild-MOG BH. Within this framework, when the energy $E$ and angular momentum $L$ of the smaller object undergo only small changes due to gravitational radiation over multiple orbital periods, the adiabatic approximation becomes applicable. As the energy and angular momentum shift, transitions in the periodic orbits can be observed, with these changes manifesting as GW. This approach allows us to track the periodic orbits dictated by the geodesic equations and calculate the gravitational radiation emitted by these orbits over time.\\
 \indent We use the kludge waveform approach outlined in \cite{babak2007kludge} to calculate GW emitted from periodic orbits around a supermassive MOG BH. In this method, the smaller object is modeled as a test particle. First, the periodic orbits are derived by solving the geodesic equations. Then, the GWs produced by these orbits are calculated by using the following formula
up to the quadratic order \cite{maselli2022detecting,liang2023probing}
\begin{align}
    h_{ij}=\frac{4\eta M}{D_L}\left(V_i V_j-\frac{m}{r}n_i n_j\right),
\end{align}
where $M$ represents the mass of the supermassive MOG BH, $m$ is the mass of the test particle, $D_L$ is the luminosity distance of the EMRI system, $\eta=\frac{Mm}{\left(M+m\right)^2}$ is the symmetric mass ratio, $V_i$ denotes the spatial velocity of the test particle, and $n_i$ is the unit vector pointing in the radial direction corresponding to the motion of the test particle. 
The GW can then be projected onto the detector-adapted coordinate system, yielding the corresponding plus $(h_+)$ and cross $(h_\times)$ polarizations, given by \cite{maselli2022detecting,liang2023probing}
\begin{align}
    h_+ &=-\frac{2\eta}{D_L}\frac{M^2}{r}\left(1+\cos^2\iota\right)\cos\left(2\phi+2\zeta\right),\\
    h_\times &=-\frac{4\eta}{D_L}\frac{M^2}{r}\cos \iota\sin\left(2\phi+2\zeta\right),
\end{align}
where $\iota$ denotes the inclination angle between the orbital angular momentum of the EMRI system and the line of sight, while $\zeta$ represents the latitudinal angle. To illustrate the GW waveform of different periodic orbits and examine how the MOG effects influence it, we consider an EMRI system consisting of a smaller component with mass $m
\ll M$ compared to the mass $M$ of the supermassive BH. For simplicity, the inclination angle $\iota$ and latitudinal angle $\zeta$ could be both set to $\frac{\pi}{4}$. Finally, since the luminosity distance, $D_L=200$ Mpc, is considered as constant in time, we have the following conclusion
\begin{align}
\label{conc1}h_+ &\propto -\frac{\cos(2\phi+2\zeta)}{r},
\end{align}
\begin{align}
     \label{conc2}h_\times &\propto -\frac{\sin(2\phi+2\zeta)}{r},
\end{align}
and we content ourselves of plotting the R.H.S's of Eqs.~\eqref{conc1} and~\eqref{conc2} versus the coordinate time $t$. In Figs.~\ref{Figwr}, \ref{Figwb} and~\ref{Figw}, we consider small values of $|\alpha|=0.05$ and we focus on the case of periodic orbits and GWs corresponding to the triplet of integers $(3, 1, 1)$.

\begin{figure}[H]
	\centering
	\begin{minipage}[b]{0.4\textwidth}
		\centering
		\includegraphics[width=\textwidth]{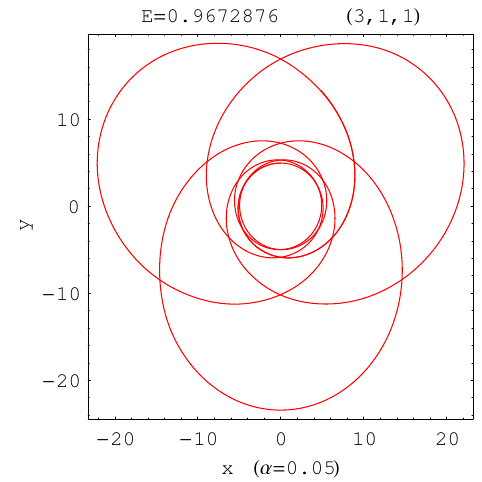}
	\end{minipage}
	\hfill
	\begin{minipage}[b]{0.5\textwidth}
		\centering
		\includegraphics[width=1\textwidth]{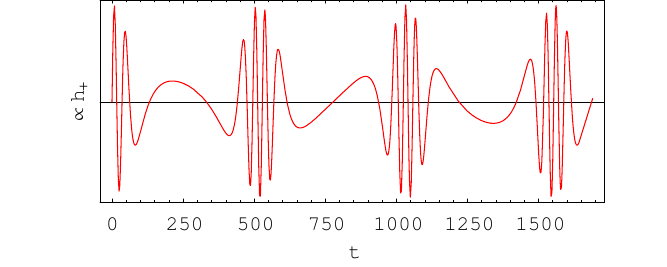}\\
		\includegraphics[width=1\textwidth]{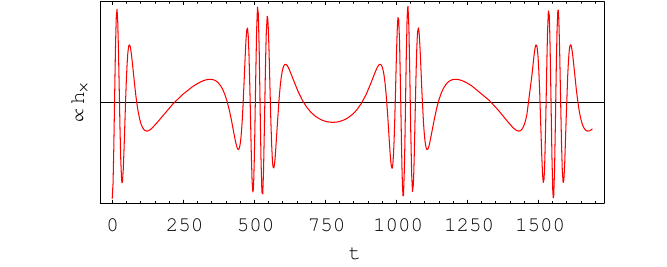}\\
	\end{minipage}
	\caption{The case $\alpha=0.05$, $L=3.8939047=L_{\text{av}}$ and $E=0.9672876$. The periodic orbit and the GW of the Schwarzschild-MOG BH for the triplet of integers $(3, 1, 1)$. In the right panels, the vertical axis represents $\cos(2\phi +2\zeta)/r$ and $\sin(2\phi +2\zeta)/r$, which are proportional to $h_+$ and $h_\times$, respectively.}
	\label{Figwr}
\end{figure}
\begin{figure}[H]
	\centering
	\begin{minipage}[b]{0.4\textwidth}
		\centering
\includegraphics[width=\textwidth]{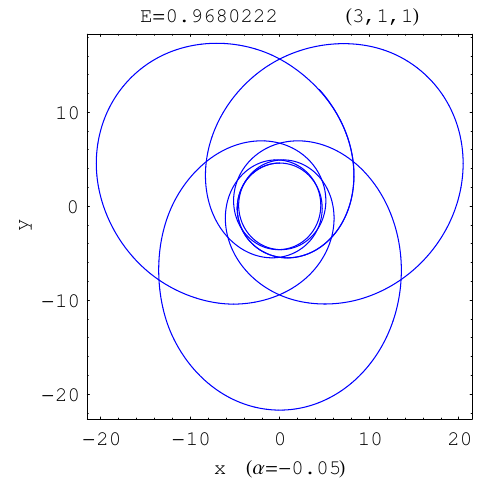}
	\end{minipage}
	\hfill
	\begin{minipage}[b]{0.5\textwidth}
		\centering
		\includegraphics[width=1\textwidth]{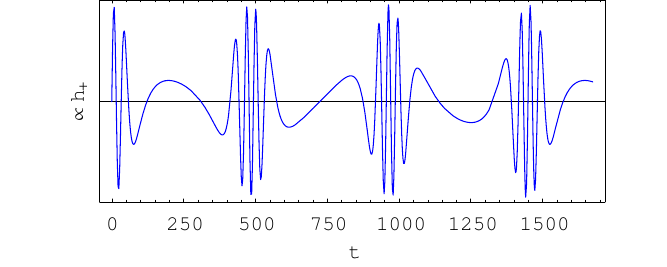}\\
		\includegraphics[width=1\textwidth]{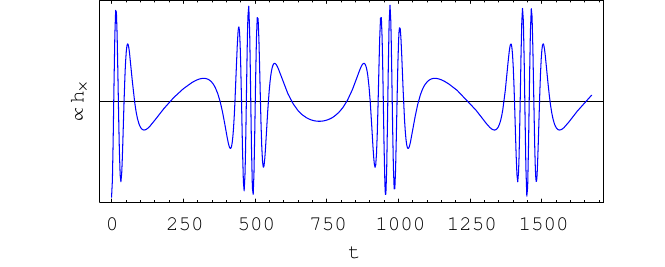}\\
	\end{minipage}
	\caption{The case $\alpha=-0.05$, $L=3.5697415=L_{\text{av}}$ and $E=0.9680222$. The periodic orbit and the GW of the Schwarzschild-MOG BH for the triplet of integers $(3, 1, 1)$. In the right panels, the vertical axis represents $\cos(2\phi +2\zeta)/r$ and $\sin(2\phi +2\zeta)/r$, which are proportional to $h_+$ and $h_\times$, respectively.}
	\label{Figwb}
\end{figure}
\begin{figure}[H]
	\centering
	\includegraphics[width=0.75\linewidth]{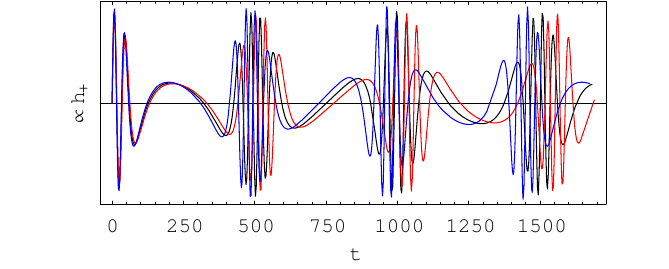}
	\includegraphics[width=0.75\linewidth]{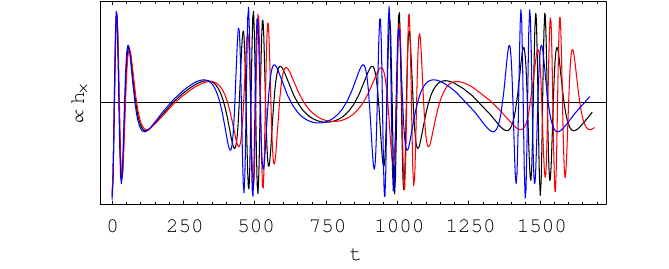}
	\caption{Upper Panel: Combination of the plots of $h_+$ in Figs.~\ref{Figwr} and~\ref{Figwb}. Lower Panel: Combination of the plots of $h_\times$ in Figs.~\ref{Figwr} and~\ref{Figwb}. In each panel, the black plot corresponds to Schwarzschild BH ($\alpha =0$) taking $L=3.7320508=L_{\text{av}}$ and $E=0.9676441$ for the same triplet of integers $(3, 1, 1)$.}
	\label{Figw}
\end{figure}
\begin{justify}
    In Figs.~\ref{Figwr2}, \ref{Figwb2} and~\ref{Figw2}, we consider small values of $|\alpha|=0.01$ and we focus on the case of periodic orbits and GWs corresponding to the triplet of integers $(4, 1, 1)$.
\end{justify}
\begin{figure}[H]
	\centering
	\begin{minipage}[b]{0.4\textwidth}
		\centering
		\includegraphics[width=\textwidth]{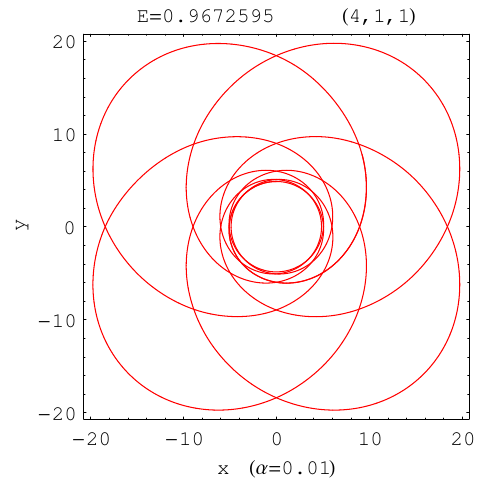}
	\end{minipage}
	\hfill
	\begin{minipage}[b]{0.5\textwidth}
		\centering
		\includegraphics[width=1\textwidth]{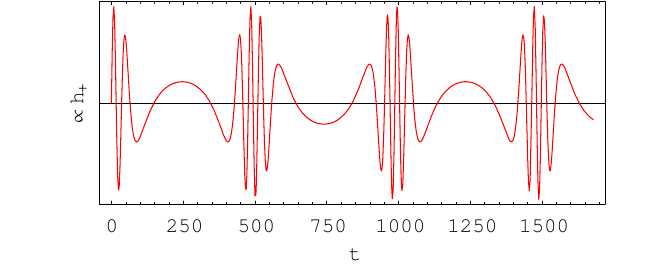}\\
		\includegraphics[width=1\textwidth]{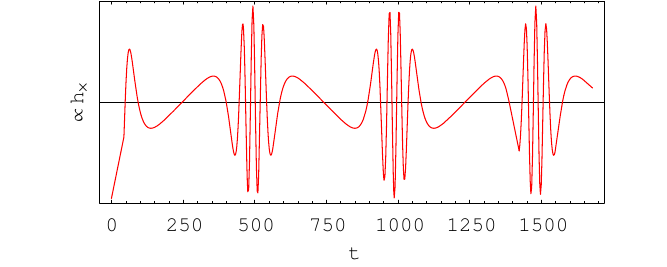}\\
	\end{minipage}
	\caption{The case $\alpha=0.01$, $L=3.7644566=L_{\text{av}}$ and $E=0.9672595$. The periodic orbit and the GW of the Schwarzschild-MOG BH for the triplet of integers $(4, 1, 1)$. In the right panels, the vertical axis represents $\cos(2\phi +2\zeta)/r$ and $\sin(2\phi +2\zeta)/r$, which are proportional to $h_+$ and $h_\times$, respectively.}
	\label{Figwr2}
\end{figure}
\begin{figure}[H]
	\centering
	\begin{minipage}[b]{0.4\textwidth}
		\centering
		\includegraphics[width=\textwidth]{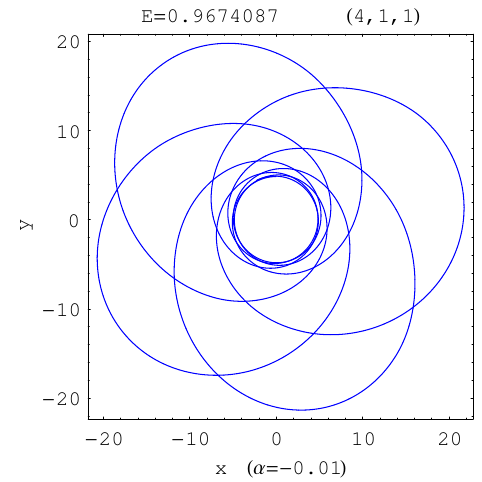}
	\end{minipage}
	\hfill
	\begin{minipage}[b]{0.5\textwidth}
		\centering
		\includegraphics[width=1\textwidth]{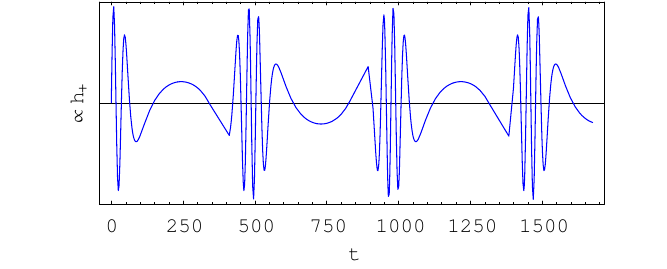}\\
		\includegraphics[width=1\textwidth]{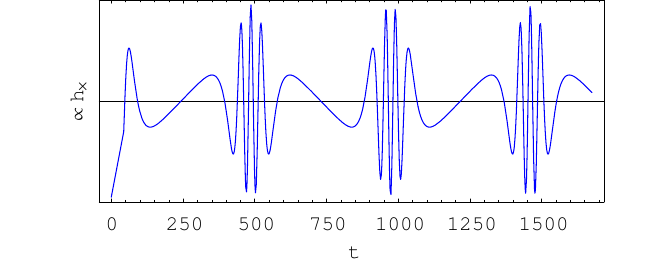}\\
	\end{minipage}
	\caption{The case $\alpha=-0.01$, $L=3.6996268=L_{\text{av}}$ and $E=0.9674087$. The periodic orbit and the GW of the Schwarzschild-MOG BH for the triplet of integers $(4, 1, 1)$. In the right panels, the vertical axis represents $\cos(2\phi +2\zeta)/r$ and $\sin(2\phi +2\zeta)/r$, which are proportional to $h_+$ and $h_\times$, respectively.}
	\label{Figwb2}
\end{figure}
\begin{figure}[H]
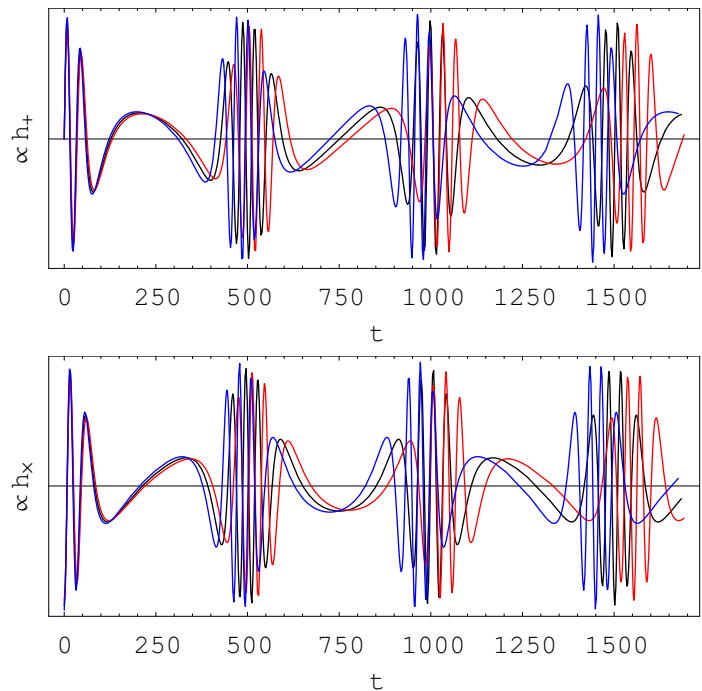

	\centering
	\includegraphics[width=0.75\linewidth]{Q311p.pdf}
	\includegraphics[width=0.75\linewidth]{Q311c.pdf}
	\caption{Upper Panel: Combination of the plots of $h_+$ in Figs.~\ref{Figwr2} and~\ref{Figwb2}. Lower Panel: Combination of the plots of $h_\times$ in Figs.~\ref{Figwr2} and~\ref{Figwb2}. In each panel, the black plot corresponds to Schwarzschild BH ($\alpha =0$) taking $L=3.7320508=L_{\text{av}}$ and $E=0.9673337$ for the same triplet of integers $(4, 1, 1)$.}
	\label{Figw2}
\end{figure}
\begin{justify}
In Figs.~\ref{Figwr}-\ref{Figw2}, the periodic orbits, characterized by a triplet of integers exhibit several zoom and whirl phases within one complete cycle. During the zoom phase, the particle follows a stretched, elliptical path, moving farther from the BH where the gravitational pull is weaker. This phase produces quieter GW signals in the $h_+$ and $h_\times$
  polarizations, aligning with the leaves of the orbit. As the particle spirals back in closer to the BH, it enters the whirl phase, tracing a tighter, circular path around the BH. This closer approach strengthens the gravitational interaction, resulting in louder, more intense bursts in the GW signal. The number of quiet phases aligns with the leaves of the orbit, while the louder bursts correspond to each whirl, reflecting the orbital structure defined by the triplet of integers.
\end{justify}
\section{Discussions and Conclusions}
This paper studies the orbital motion of timelike geodesics around the Schwarzschild-MOG BH, considering the MOG parameter $\alpha$. We derived the geodesic equations using the Hamiltonian formulation and explored how different values of $\alpha$ affect the spacetime structure, particularly the locations of the event and Cauchy horizons. We find that as $\alpha$ increases, the horizons move farther apart, and for $\alpha=-1$, the BH reaches the extremal case. Our analysis includes the investigation of effective potentials, ISCO, and IBCO. It is also noted that the effective potential tends to $1$ as $r\rightarrow\infty$, leading to an asymptotically flat spacetime.\\
\indent To reduce the number of free parameters, we defined a new angular momentum, $L_{\text{av}}$, which is the average of the angular momenta of the ISCO and IBCO for any value of $\alpha$. By plotting the effective potential against $r$ for these specific angular momenta, we found that for $0\leq \alpha <\infty$, a deep potential well forms, which becomes shallower and smaller for $-1<\alpha <0$. This well allows for a diverse range of periodic orbits, leading us to focus on the range $\alpha \in(-1,\infty)$ for periodic orbit analysis. 
We introduced the rational number $q$, along with a triplet of integers $(z,w,v)$, to classify the orbits. Energy tables for different values of $q$ were constructed, and a series of periodic orbits were obtained, as shown in Figs.~\ref{fig 6} and~\ref{fig 7}. Our analysis of orbital motion between the ISCO and IBCO reveals that for $\alpha>0$, the energy required for periodic orbits in Schwarzschild-MOG BH is lower than in Schwarzschild BH, while for $-1<\alpha<0$, the energy values are higher.\\ \indent Furthermore, by considering the effect of an irrational number, we observed orbital precession due to small perturbations and plotted corresponding graphs to demonstrate this phenomenon. Finally, we examined GW radiations emitted by periodic orbits characterized by different values of $(z,w,v)$ in the Schwarzschild-MOG BH. The analysis reveals distinct zoom-whirl patterns, as shown in Figs.~\ref{Figwr}-\ref{Figw2}. The GW waveforms $h_+$
  and $h_\times$ demonstrate distinct quiet phases during the highly elliptical zooms, transitioning to louder glitches during the nearly circular whirls. The quiet phases correspond to the leaves of the orbits, while the louder glitches match the number of whirls. These patterns reveal the influence of the MOG parameter $\alpha$ on the phases of the GW signals, providing key insights into the orbital dynamics of EMRI systems. Moreover, these findings offer observational signatures that could help constrain the Schwarzschild-MOG BH geometry through future GW detectors.\\ \indent In comparison to the study on periodic orbits around polymer BH in loop quantum gravity (LQG) \cite{tu2023periodic}, both works examine the effects of MOG on orbital dynamics and GW emissions. Our study focuses on the Schwarzschild-MOG BH, introducing the parameter $\alpha$ to modify gravitational interactions, while the LQG study investigates polymer BHs. Both studies observe zoom-whirl behavior in periodic orbits and distinct quiet and glitch phases in GW signals. Although the LQG study highlights the impact of the polymer model, our work emphasizes how the MOG parameter affects spacetime structure, effective potentials, and orbital energy. Both studies suggest that GW signals can distinguish these modified gravity models from traditional Schwarzschild BHs, providing key insights for future observations.\\
  \indent In future research, it would be compelling to extend this framework to study periodic orbits and GW signals in other alternative theories of gravity. Additionally, exploring more compact objects, such as boson stars or exotic BH solutions, could offer new perspectives on the interplay between orbital dynamics and GW emissions. Such investigations could lay the foundation for developing more precise tools to test MOG theories and enhance our understanding of the universe through upcoming GW observations.
\end{justify}
 \bibliography{reference}
\bibliographystyle{ieeetr}
\end{document}